\documentclass[runningheads]{llncs}
\usepackage{graphicx}
\usepackage{caption}
\usepackage{footnote}
\usepackage{comment}
\usepackage{mathtools}
\usepackage{afterpage}
\usepackage{multirow}

\begin{document}
\title{Twenty Years of Network Science: \\ A Bibliographic and Co-Authorship \\ Network Analysis}
\titlerunning{Twenty Years of Network Science}
% If the paper title is too long for the running head, you can set
% an abbreviated paper title here
%

\author{Roland Molontay\inst{1,2,3}\orcidID{0000-0002-0666-5279}\and Marcell~Nagy\inst{1,2}\orcidID{0000-0001-5666-7777} 
}
\authorrunning{R. Molontay and M. Nagy}
% First names are abbreviated in the running head.
% If there are more than two authors, 'et al.' is used.
%
\institute{Dept. of Stochastics, Budapest University of Technology and Economics, Hungary 
\and
Faculty of Informatics, University of Debrecen, Hungary
\and
MTA--BME Stochastics Research Group, Hungary \\
\email{\{molontay, marcessz\}@math.bme.hu}}
\maketitle              % typeset the header of the contribution
\begin{abstract}%250 szó
Two decades ago three pioneering papers turned the attention to complex networks and initiated a new era of research, establishing an interdisciplinary field called network science. Namely, these highly-cited seminal papers were written by Watts~\&~Strogatz,  Barab\'asi~\&~Albert, and Girvan~\&~Newman on small-world networks, on scale-free networks and on the community structure of complex networks, respectively. In the past 20 years -- due to the multidisciplinary nature of the field -- a diverse but not divided network science community has emerged. In this paper, we investigate how this community has evolved over time with respect to speed, diversity and interdisciplinary nature as seen through the growing co-authorship network of network scientists (here the notion refers to a scholar with at least one paper citing at least one of the three aforementioned milestone papers). After providing a bibliographic analysis of 31,763 network science papers, we construct the co-authorship network of 56,646 network scientists and we analyze its topology and dynamics. We shed light on the collaboration patterns of the last 20 years of network science by investigating numerous structural properties of the co-authorship network and by using enhanced data visualization techniques. We also identify the most central authors, the largest communities, investigate the spatiotemporal changes, and compare the properties of the network to scientometric indicators.

%This paper honors the contributions of network science by exploring the evolution of this community as seen through the growing co-authorship network of network scientists (here the notion refers to a scholar with at least one paper citing at least one of the three aforementioned milestone papers). 

\end{abstract}
%
%
%
% \section{First Section}
% \subsection{A Subsection Sample}
% Please note that the first paragraph of a section or subsection is
% not indented. The first paragraph that follows a table, figure,
% equation etc. does not need an indent, either.

% Subsequent paragraphs, however, are indented.
\section{Introduction}
Complex networks have been studied extensively since they efficiently describe a wide range of systems, spanning many different disciplines, such as Biology (e.g. protein interaction networks), Information Technology (e.g., WWW, Internet), Social Sciences (e.g., collaboration, communication, economic, and political networks), etc.  Moreover, not only the networks originate from different domains, but the methodologies of network science as well, for instance, it heavily relies on the theories and methods of graph theory, statistical physics, computer science, statistics, and sociology.

In the last two decades, network science has become a new discipline of great importance. It can be regarded as a new academic field since 2005 when the U.S. National Research Council defined network science as a new field of basic research~\cite{national2005Network}. The most distinguished academic publishing companies announce the launch of new journals devoted to complex networks, one after another (e.g. Journal of Complex Networks by Oxford University Press, Network Science by Cambridge University Press, Applied Network Science and Social Network Analysis and Mining by Springer). Network science also has its own prestigious conferences attended by thousands of scientists. Leading universities continuously establish research centers and new departments for network science, furthermore, launch Master and Ph.D. programs in this field (such as Yale University, Duke University, Northeastern University, and Central European University).

The significance of network theory is also reflected in the large number of publications on complex networks and in the enormous number of citations of the pioneering papers by Barab\'asi~\&~Albert~\cite{barabasi1999emergence}, Watts~\&~Strogatz~\cite{watts1998collective} and Girvan~\&~Newman~\cite{girvan2002community}.  Some researchers interpret network science as a new paradigm shift~\cite{kocarev2010network}. However, complex networks are not only acknowledged by the research community, but innovative textbooks aimed for a wider audience have also been published~\cite{barabasi2016network,newman2018networks}, moreover, the concepts of network science have appeared in the popular literature~\cite{barabasi2003linked,watts2004six} and mass media~\cite{connected2008} as well.

In the last two decades, complex networks became in the center of research interest thanks to --~among many others~-- the aforementioned three pioneering papers and due to the fact that the prompt evolution of information technology has opened up new approaches to the investigation of large networks. This period of 20 years can be regarded as the golden age of network science. The first challenge was to understand network topology, to this end, structural properties were put under the microscope one after the other (small-worldness, scale-free property, modularity, fractality, etc.) and various network models were proposed to understand and to mathematically describe the architecture and evolution of real-world networks~\cite{vespignani2018twenty}. In recent years, there has been a shift from the structural analysis to studying the control principles of complex networks~\cite{barabasi2019twenty}.  Remarkable computing power, massive datasets, and novel computational techniques keep great potential for network scientists for yet another 20 years~\cite{vespignani2018twenty}.

This work is a tribute to the achievements of the network science community in the past 20 years. We provide a bibliographic analysis of 31,763 network science papers and  we also construct and investigate the co-authorship network of network scientists to identify how the network science community has been evolving over time.

The present study also extends the earlier conference version of this paper~\cite{molontay2019two} in several important directions. Namely, here we provide a more detailed literature review; we examine a longer time period; and we answer the question of how the network science community has evolved over time with respect to speed, diversity, and interdisciplinary nature by implementing novel analyses. Here we also provide an analysis of the co-occurrence network of the keywords. Moreover, we supplement our previous work with several other new methods and data visualizations that help to make insightful observations regarding the last two decades of network science.

The main contributions of this work can be summarized as follows:
\begin{itemize}
    \item We collect 31,763 network science papers and provide a bibliographic analysis investigating various characteristics of the papers and showing how the discipline has developed over time.
    \item We construct the co-authorship network of 56,646 network scientists and undertake a scholarly network analysis study by analyzing its topology and dynamics.
    \item We answer the following major research questions:
    \begin{itemize}
        \item What are the most important venues of network science and how have they changed over time?
        \item How the publication patterns vary over research areas and time?
        \item What are the most important topics of network science and how have they evolved through time? What relationships can we explore among the most frequent keywords of network science?
        \item How the network science community has evolved over time with respect to speed, diversity, and interdisciplinary nature?
        \item What are the most typical patterns in terms of international and interdisciplinary collaborations?
        \item Who are the most central authors and how do the largest communities look like? How do these network properties compare to other scientometric indicators?
    \end{itemize}
\end{itemize}

\section{Scholarly networks analysis}
The present paper joins the line of research focused on scholarly network analysis that is based on big scholarly data \cite{pawar2019codd,yan2014scholarly}. Big scholarly data refers to the rapidly growing data accessible in digital libraries containing information on millions of scientific publications and authors \cite{xia2017big}.
The easily available data sources (Web of Science, Scopus, PubMed, Google Scholar, Microsoft Academic, the U.S. Patent and Trademark Office, etc.) together with novel powerful data analysis technologies have led to the emergence of science of science \cite{fortunato2018science} that gives us a better understanding of the self-organizing rules and patterns of science, e.g. how disciplines have emerged and evolved over time \cite{leonidou2010five}. Various scholarly networks at many levels can be formed based on scholarly data; Pawar \textit{et al.} identify the following forms of scholarly networks of great interest \cite{pawar2019codd}:
\begin{enumerate}
    \item co-authorship networks (a link is formed between scientists by their co-authorship of at least one scientific paper),
    \item citation networks (a directed link is formed between documents referencing one another),
    \item co-citation networks (a link is formed between documents if they are cited together),
    \item bibliographic coupling (documents are linked if they share common references),
    \item co-occurrence networks (keywords/topics are linked if they occur in the same document), and
    \item heterogeneous networks (two or more coupled scholarly networks).
\end{enumerate}

Among the aforementioned scholarly networks perhaps co-authorship networks have attracted the greatest deal of research interest, owing to the fact that co-authorship is one of the most important reflections of research collaboration, which is an essential mechanism that joins together distributed knowledge and expertise into novel discoveries. Furthermore, building the map of sciences is not only important for sociologists and other scholars to understand researchers' interaction but for policymakers as well to address sharing resources \cite{xia2017big}. Co-authorship networks have been studied extensively in various ways and from various perspectives: e.g. the collaboration network determined by the articles of a certain journal, a specific country or a research community that cites a particular influential paper or author~\cite{barabas2019co,barabas2017impact,barabasi2002evolution,kumar2015co,newman2001structure,newman2004coauthorship}.  In this paper, we investigate the co-authorship network of network scientists as defined in the following section.

Keywords co-occurrence networks have also been investigated thoroughly \cite{li2016evolutionary,su2010mapping,uddin2015framework}. Keywords of academic articles can provide a concise overview of the content and the core idea of the body of the papers. In contrast to word clouds, co-occurrence networks do not only show the frequency of the keywords but also allow us to discover the relationship between them. Li \textit{et al.} investigated  6,900 articles published between 1982 and 2013 which had been indexed using the keyword 'complex network(s)' and provided a co-keyword network and a keyword co-occurrence network analysis \cite{li2016evolutionary}.

\section{Preliminaries and data}
In this section, we describe how the co-authorship network of network scientists was constructed, how the examined set of academic publications was chosen and collected, what data preparation steps were conducted. Moreover, we also present some useful notions and preliminaries.

\subsection{Co-authorship network of network scientists}
To the best of our knowledge, the co-authorship network of network scientists has been analyzed only by Newman \textit{et al.}~\cite{newman2006finding,newman2004finding}. However, their network consists of 1,589 authors, while this study investigates a much larger network (56,646 vertices and 357,585 edges) spanning a longer time horizon (1998-2019).  

We construct the co-authorship network of network scientists as follows. We consider three ground-breaking papers around the millennium that can be regarded as the roots of the rise of network science: the paper of Watts~\&~Strogatz~\cite{watts1998collective} on small-world networks, the work of Barab\'asi~\&~Albert~\cite{barabasi1999emergence} about scale-free networks and the paper of Girvan~\&~Newman~\cite{girvan2002community} that reveals the community structure of complex networks. We selected the aforementioned three papers since they initiated new areas of research in network science by introducing pivotal concepts two decades ago, that had a huge impact on the network science community that is also demonstrated by the large number of citations they received in the past 20 years.

\begin{figure}[h]
\vspace{-12pt}
    \centering
    \begin{minipage}{0.48\textwidth}
        \centering
         \includegraphics[width=\linewidth]{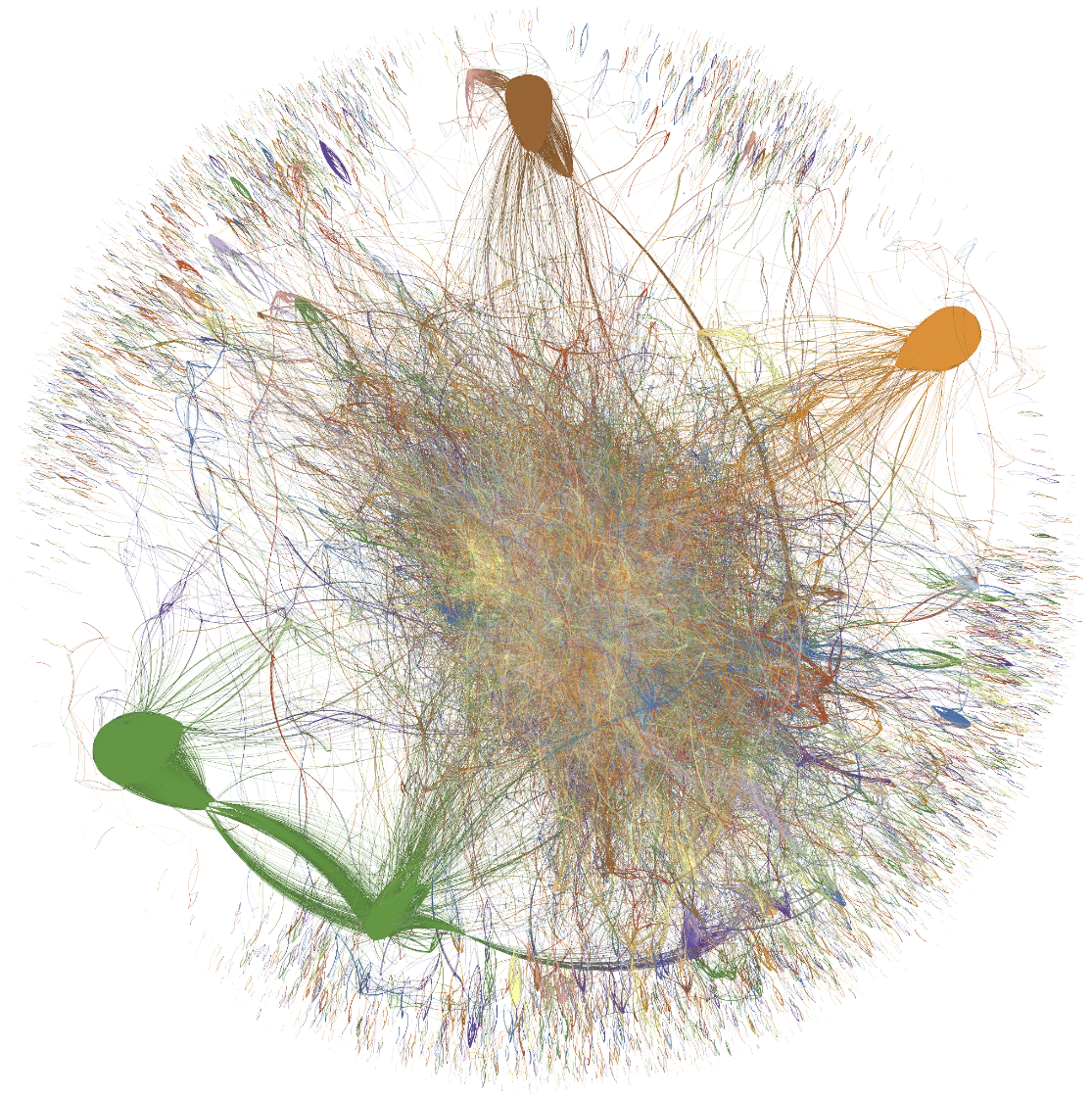}
    \caption{The co-authorship network of network scientists colored by communities.}
    \label{fig:NNS}
    \end{minipage}%
    \hfill
    \begin{minipage}{0.48\textwidth}
        \centering
      \includegraphics[width=\linewidth]{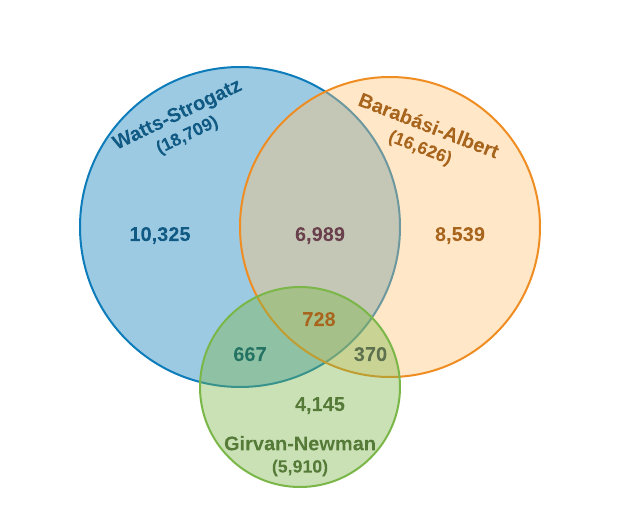}
    % \vspace{-5pt}
    \caption{Distribution of the citations among the three pioneering papers.}
    %\vspace{-10pt}
    \label{fig:venn}
    \end{minipage}
\end{figure}

In this work, we consider a paper as a \textit{network science paper} if it cites at least one of the three aforementioned pivotal articles (in addition, the three originating papers are also regarded as network science papers, obviously). Similarly, we call someone a \textit{network scientist} if (s)he has at least one network science paper. The previous definitions of a network science paper and a network scientist are of course quite arbitrary. It is important to note that the papers that refer to one of the three seminal papers are not necessarily about network science and there certainly exist network science articles that do not refer to any of the aforementioned pioneering papers. On the other hand, we believe that this concept is a good proxy for our purposes and it is worth studying. We construct the co-authorship network of the network scientists where two of them are connected if they have at least one joint network science paper (see Fig.~\ref{fig:NNS}). In other words, this network is a one-mode projection onto scientists, from the bipartite network of scientists and the network science papers they authored.  The anonymized data of the constructed network and some figures in high resolution are available in the supplementary material~\cite{supp}.

\vspace{-20pt}
\subsection{Glossary}

\begin{definition}[Graph]
A simple (undirected) graph is an ordered pair $G=(V,E)$, where $V$ is the set of vertices or nodes and $E$ is the set of edges or links, which are two-element subsets of $V$. The vertex and edge sets of $G$ are denoted by $V(G)$ and $E(G)$ respectively. The size of the graph is the number of its nodes, and it is usually denoted by $n$. 
\end{definition}

\begin{definition}[Complex network]
In network theory, the terms graph and network are used interchangeably, however, a complex network is a graph with non-trivial topological features, that characterize real-world networks.
\end{definition}

\begin{definition}[Average path length]
A path is a sequence of edges which connect a sequence of vertices. The distance $d(u,v)$ between the vertices $u$ and $v$ is the length (number of edges) of the shortest path connecting them. The $l_G$ average path length of a graph $G$ of size $n$ is defined as:
$$l_G = \frac{1}{n(n-1)}\sum_{\substack{u,v \in V(G) \\ u\neq v}} d(u,v). $$
\end{definition}

\begin{definition}[Small-world property]
A network is said to be small-world, if the average path length is proportional to the logarithm of the size of the network i.e. $l_G \sim \log |V|$.
\end{definition}

\begin{definition}[Degree distribution]
The degree $\deg(v)$ of a vertex $v$ in a simple, undirected graph is its number of incident edges. The degree distribution $P$ is the probability distribution of the degrees over the whole network, i.e. $P(k)$ is the probability that the degree of a randomly chosen vertex is equal to $k$.
\end{definition}

\begin{definition}[Scale-free property]
A scale-free network is a connected graph which $P(k)$ degree distribution follows a power law asymptotically, i.e. $P(k) \sim k^{-\gamma},$ where $\gamma \geq 1$.
\end{definition}

\begin{definition}[Assortativity coefficient]
The assortativity coefficient is the Pearson correlation coefficient of degree between pairs of linked nodes. The assortativity coefficient is given by  $$r=\frac{\sum_{j,k}{j\cdot k (e_{j,k}-q_{j}q_{k})}}{\sigma_ q^2},$$ where the term $q_{k}$ is the mass function of the distribution of the remaining degrees (degree of the nodes minus one) and $j$ and $k$ indicates the remaining degrees. Furthermore, $e_{j,k}$ refers to the mass function of the joint probability distribution of the remaining degrees of the two vertices. 
Finally, $\sigma_q^2$ denotes the variance of the remaining degree distribution with mass function $q_k$  i.e. $\sigma_q^2 = \sum_k k^2q_k - \left(\sum_k k q_k \right)^2.$
\end{definition}

\begin{definition}[Local clustering coefficient]
The local clustering coefficient of vertex $v$ is the fraction of pairs of neighbors of $v$ that are connected over all pairs of neighbors of $v$. Formally:
$$C_{\mathrm{loc}}(v) = \frac{| \{(s,t) \text{ edges}: s,t \in N_v \text{ and } (s,t) \in E|\}}{\deg(v) (\deg(v)-1)},$$
where $N_v$ is the neighborhood  of the node $v$ i.e. the vertices adjacent to $v$.

The average (local) clustering coefficient of a $G$ graph is defined as:
$$\Bar{C}(G) = \frac{1}{n}\sum_{v\in V(G)} C_{\mathrm{loc}}(v), $$ where $n$ is the size of the graph.
\end{definition}

\begin{definition}[Global clustering coefficient]
The global clustering coefficient $C$ of the graph G is the fraction of closed triplets (paths of length two in G that are closed) over all of the triplets (paths of length two) in $G$.
\end{definition}

\begin{definition}[Betweenness centrality]
Betweenness centrality of a node $v$ is given by the expression:
$$ g(v) = \sum_{s \neq v \neq t} \frac{\sigma_{st}(v)}{\sigma_{st}}, $$ where $\sigma_{st}$ is the total number of shortest paths from node $s$ to node $t$ and $\sigma_{st}(v)$ is the number of those paths that pass through $v$.
\end{definition}

\begin{definition}[$h$-index]
The $h$-index is an author-level metric defined as the maximum value of $h$ such that the given author has published $h$ papers that have each been cited at least $h$ times.
\end{definition}

\begin{definition}[Harmonic centrality] Harmonic centrality of a node $v$ is the sum of the reciprocal of the shortest path distances from all other nodes to $v$:

$$
H(v) = \sum_{u\neq v}\frac{1}{d(u,v)}.
$$

\end{definition}

\subsection{Data collection and preparation}

We build our analysis on data collected from the Web of Science bibliographic database, retrieved on January 2, 2020. The collected data consist of 41,245 rows with multiplicity corresponding to the citing works of the three seminal articles~\cite{barabasi1999emergence,girvan2002community,watts1998collective}. For each citing paper we have information on the document title, publication name, publication type, publisher, publication year, authors' full name, authors' address,  research area, keywords, cited reference count, total times cited count, page count, abstract, etc.

After the data were collected, various data preparation steps were conducted, including merging the files,  handling missing fields, deleting duplicates, and indicating which of the three seminal papers were cited by the given article. These preparation steps reduced the dataset to 31,763 unique rows and the citation pattern of the corresponding articles is shown in Fig.~\ref{fig:venn}.

The authors are represented by the full name field of Web of Science, however, this field is unfortunately not consistent, the author called John Michael Doe may appear as Doe,~John; John Doe; Doe, J.; Doe, J. M.; Doe, John Michael, and other variants. To overcome this issue, we created a dictionary that defines the name variants that correspond to the same author. Furthermore, we cannot distinguish between different scientists with the same name, this issue is mainly relevant for Asian authors. However, the error introduced by this problem is negligible, as also pointed out by Newman~\cite{newman2001structure} and by Barab\'asi \textit{et al.}~\cite{barabasi2002evolution}.

\section{Analysis of network science papers}

\begin{figure}[h]
    \centering
    \includegraphics[width=0.72\linewidth]{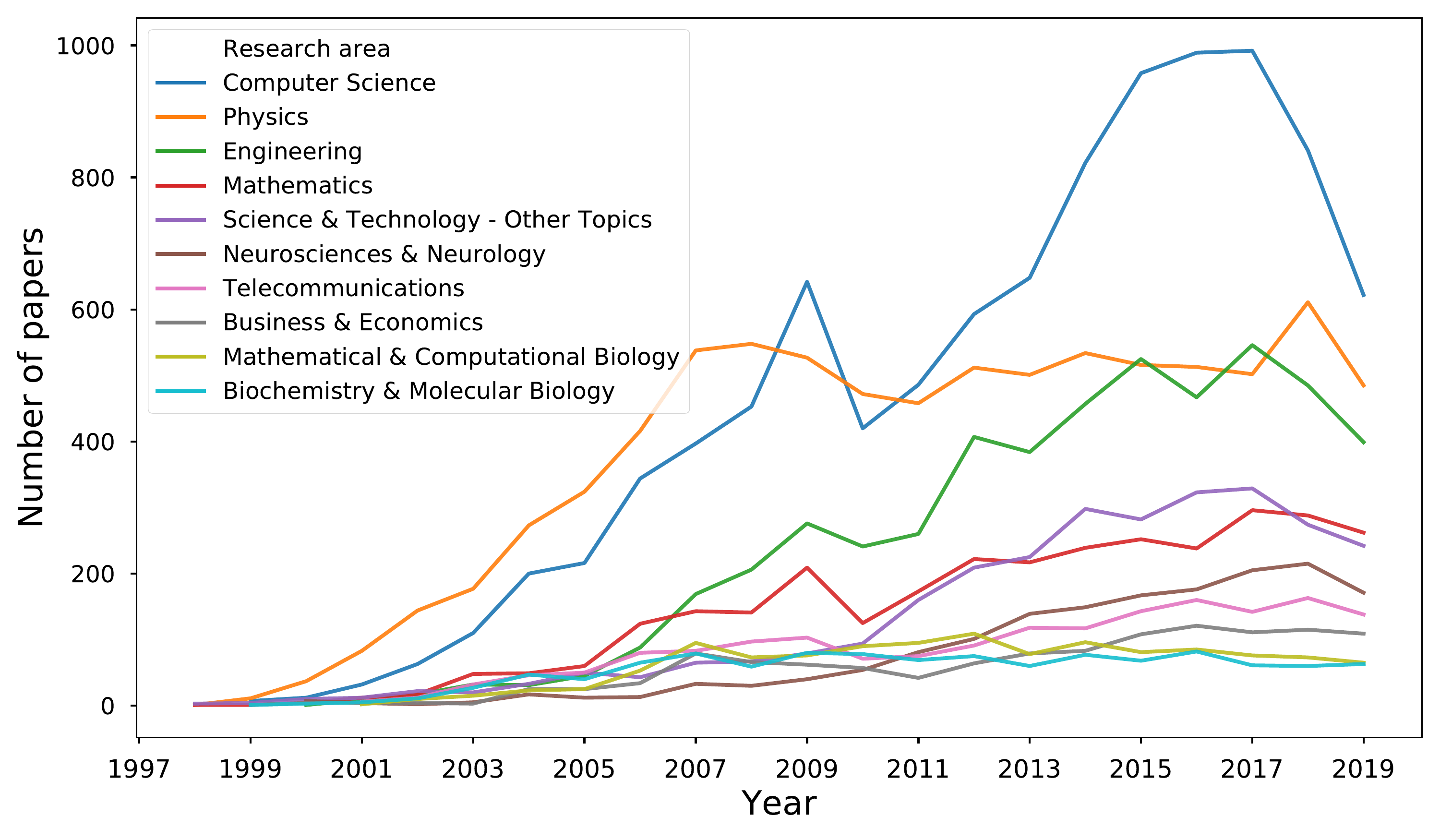} 
    \caption{Top 10 research areas of the network science papers}
    \label{fig:ra_linechart}
    \vspace{-12pt}
\end{figure}

\begin{figure}[h]
    \centering
    \includegraphics[width=0.72\linewidth]{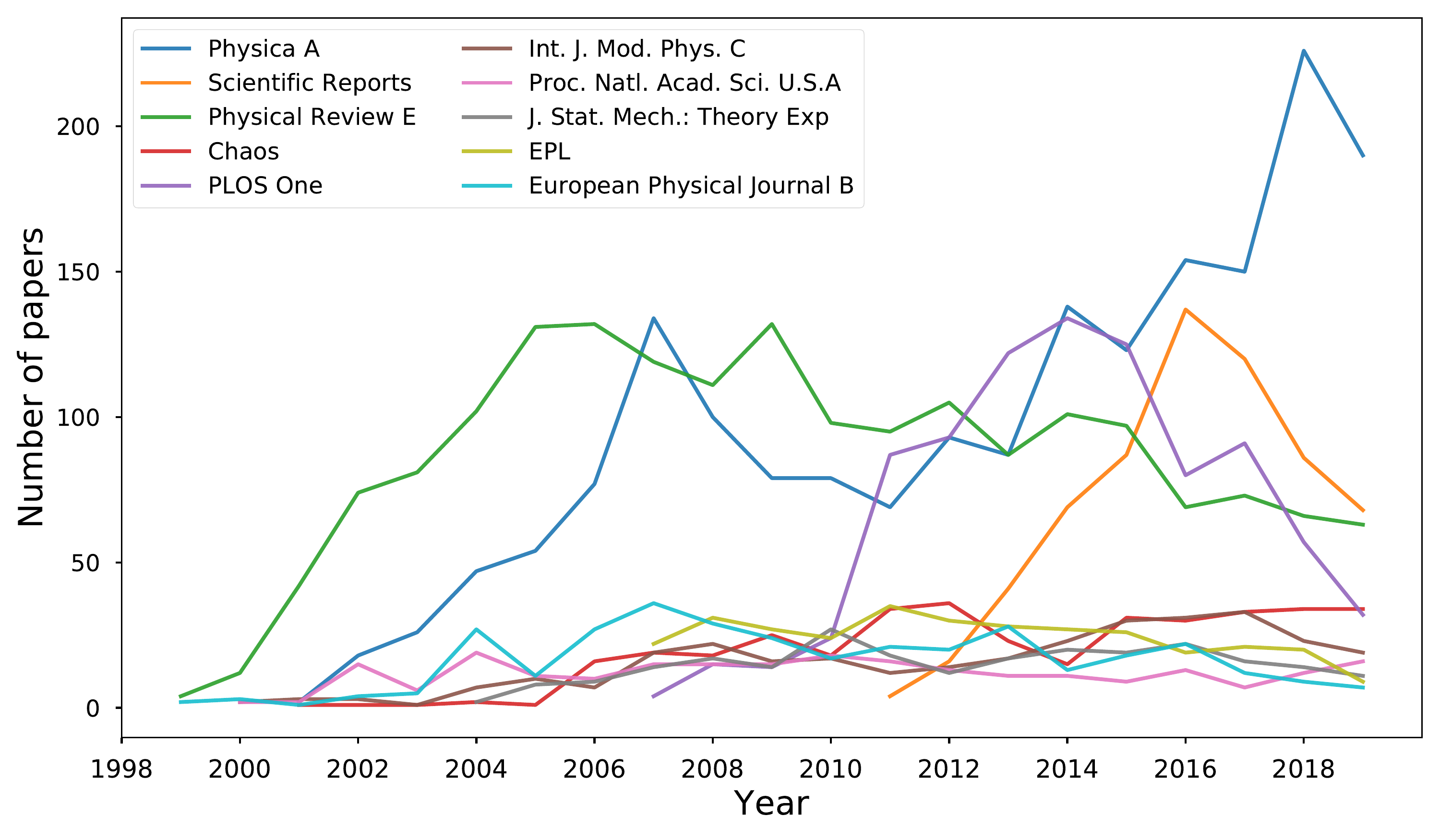}
    \caption{Top 10 journals of the network science papers}
    \label{fig:journal_linechart}
    \vspace{-12pt}
\end{figure}

First, we analyze the enormous number of citing works, i.e. the network science papers. Fig.~\ref{fig:ra_linechart} shows the top 10 research areas where the citing works belong to, illustrating the interdisciplinary nature of network science. We can see that the first decade was dominated by physics while later computer science took over. It is also clear from the figure that neuroscience has started to use tools of network science in the last decade. The journals that publish the most network science papers are shown in Fig.~\ref{fig:journal_linechart}. Considering the number of publications, Physical Review E was the leading scientific forum of network science in the first half of the examined period, while PLOS One and Scientific Reports emerged in the last decade. Currently, Physica A can be regarded as the leading journal of network science in terms of the number of published network science articles.


\begin{figure}
    \centering
    \includegraphics[width=0.6\linewidth]{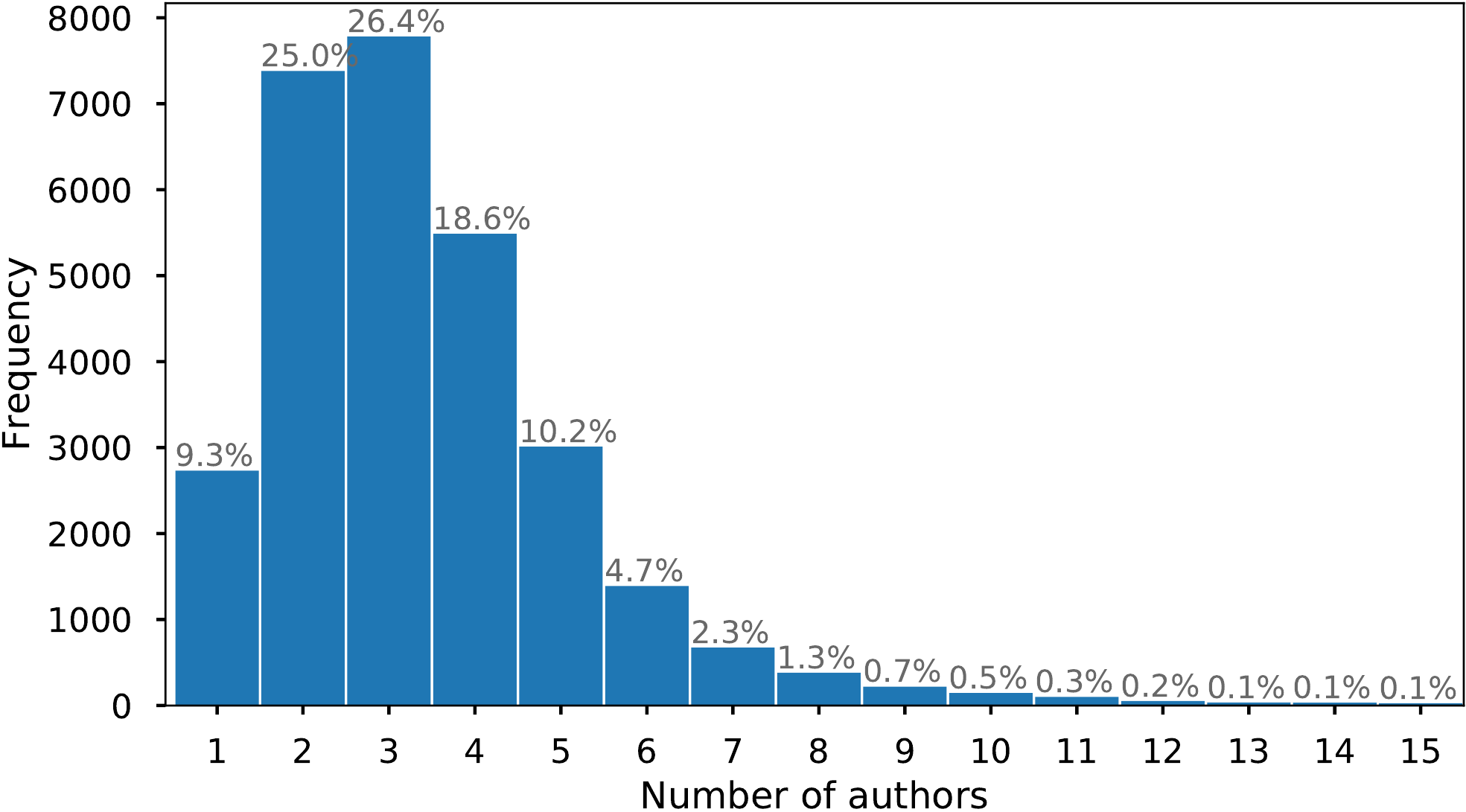}
    %\vspace{-9pt}
    \caption{Histogram of the number of authors per paper (truncated at 15).}
    \label{fig:authors_per_paper}
    \vspace{-12pt}
\end{figure}

Fig.~\ref{fig:authors_per_paper} shows the number of collaborating authors per citing works, the most typical numbers of co-authors in a network science paper are 2 and 3. Almost one-tenth of the network science papers are sole-authored, M. E. J. Newman has the highest number of sole-authored network science papers, namely 27.  While the figure shows only up to 15 number of authors, there are a few papers with a high number of collaborating authors e.g. the paper with the highest number of authors (387) is a paper of the Alzheimer’s Disease Neuroimaging Initiative~\cite{lella2018communicability}. The authors of this article emerge as a maximal clique of the co-authorship network of network scientists as it can be seen in Fig.~\ref{fig:NNS}.

We also investigate the distribution of network science papers written by network scientists. The authors with the highest number of network science papers together with the citation count of their network science papers are shown in Table~\ref{top10}. Guanrong Chen has the highest number of network science papers, his research areas are nonlinear systems and complex network dynamics and control.

\begin{figure}
    \centering
    \includegraphics[width=0.7\linewidth]{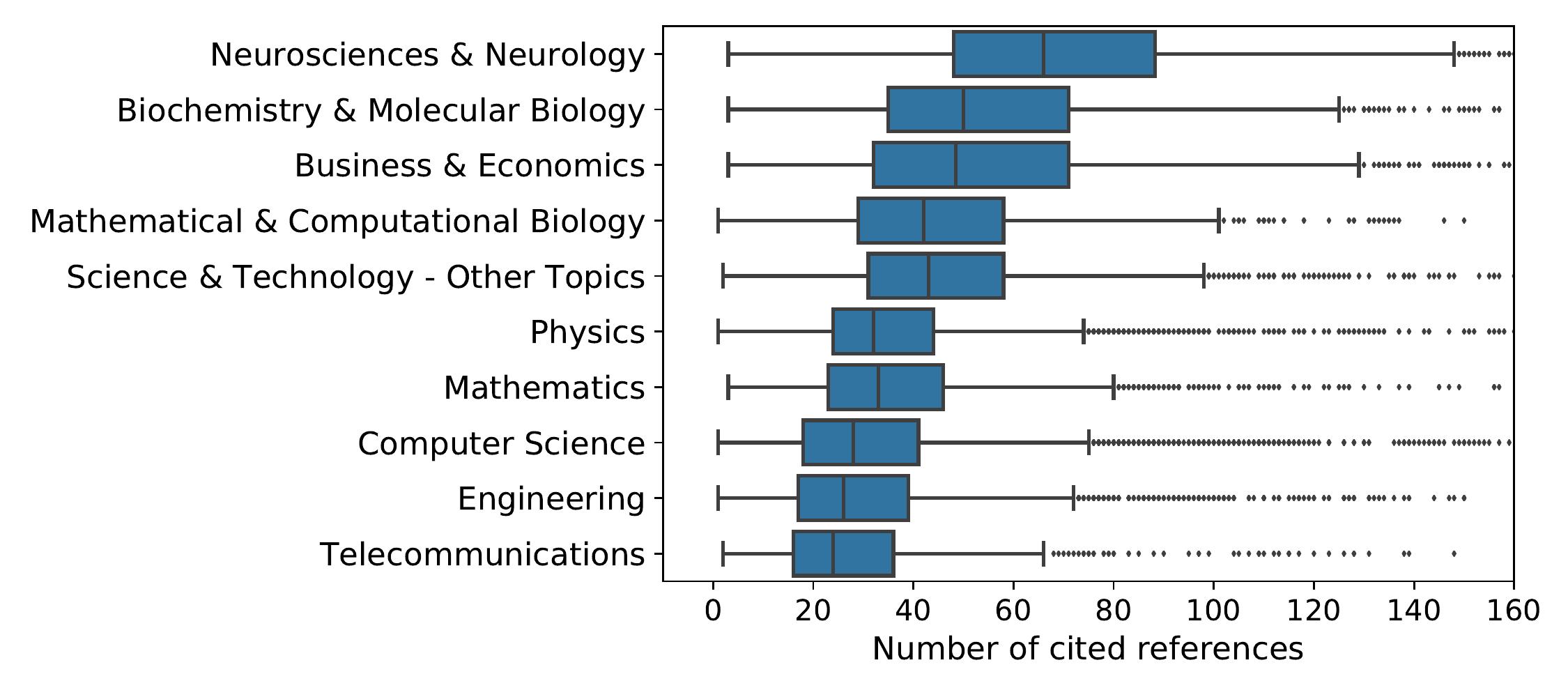}
     \includegraphics[width=0.7\linewidth]{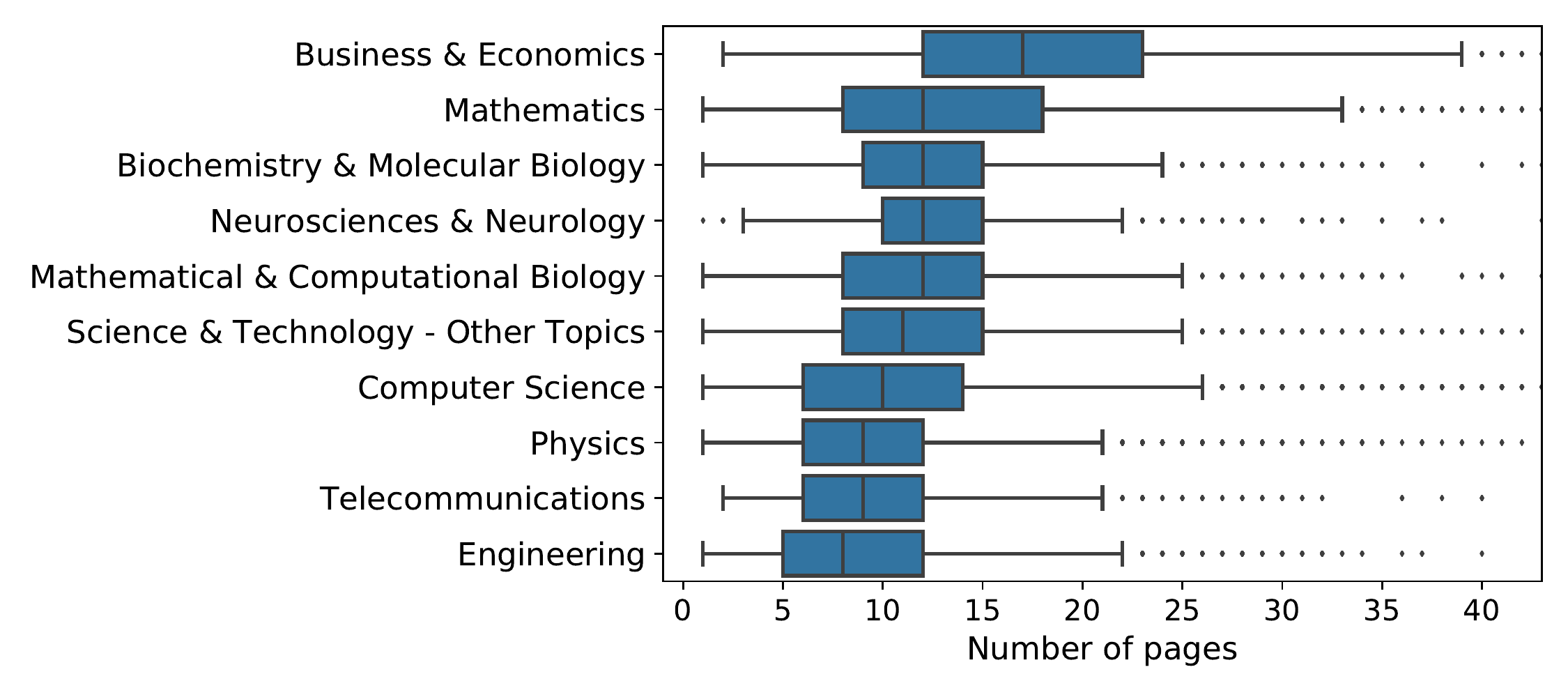}
    \caption{Boxplots of the number of cited references and length of network science papers across research areas.}
    \label{fig:cited_references}
    \vspace{-12pt}
\end{figure}

\begin{table}[h]
% \vspace{-6pt}
\caption{Top 12 authors with the most network science papers.}
\label{top10}
\centering
\begin{tabular}{lcc}
\textbf{Name of author} & \multicolumn{1}{l}{\textbf{Number of papers}} & \multicolumn{1}{l}{\textbf{Number of citations}} \\ \hline
Guanrong Chen & 167 & 12,859\\
Bing-Hong Wang & 145 &  5,341\\
Tao Zhou & 138 &  9,911\\
Shlomo Havlin & 124 & 13,377\\
J\"urgen Kurths & 118 & 9,249\\
Eugene H. Stanley & 113 &  10,479\\
Zhongzhi Zhang & 104 &  2,099\\
Ying-Cheng Lai & 99 &  5,799\\
Albert-L\'aszl\'o Barab\'asi & 95 &  73,937\\
Luciano da Fontoura Costa & 92 & 2,726\\
Matjaz Perc & 89 & 8,634\\
Michael Small & 88 & 2,345
\end{tabular}
\vspace{-12pt}
\end{table}
To gain some insights on how the publication patterns vary in network science depending on the research area, we show how the distribution of the number of cited references and the length of the papers differ across research areas (see Fig.~\ref{fig:cited_references}). We can observe that in neuroscience and neurology authors typically cite a high number of articles while in computer science or engineering the typical number of cited references is much smaller. Engineering together with telecommunications and physics are also the areas with the shortest articles, while it can also be due to the fact that in those disciplines double-column publication formats are quite typical. It is important to emphasize that these observations are not necessarily representative of the disciplines in general, only for those papers that were defined as network science papers.

\begin{figure}[h]
    \centering
    {\includegraphics[width=0.48\linewidth]{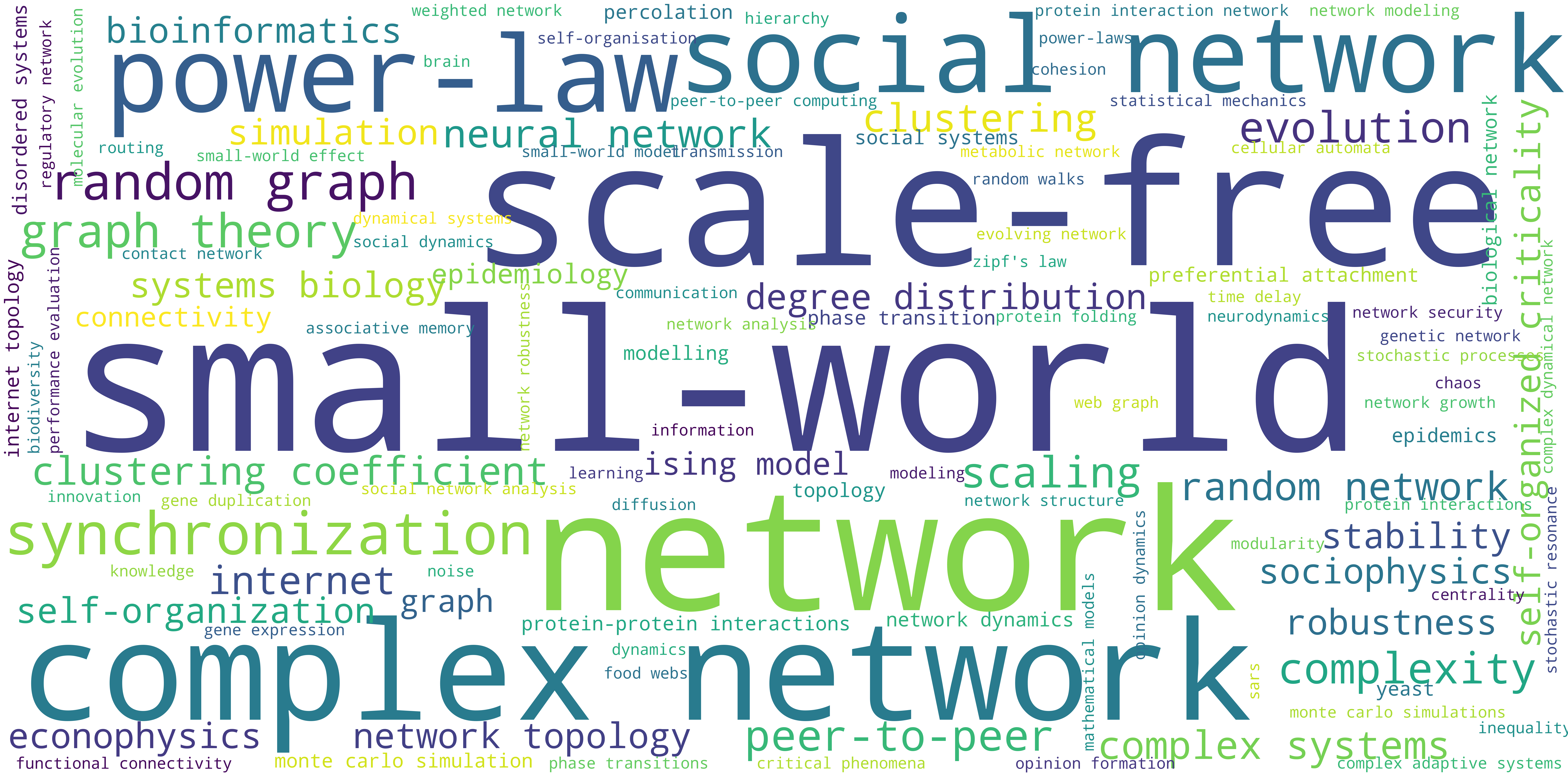} 
    \hfill
    \includegraphics[width=0.48\linewidth]{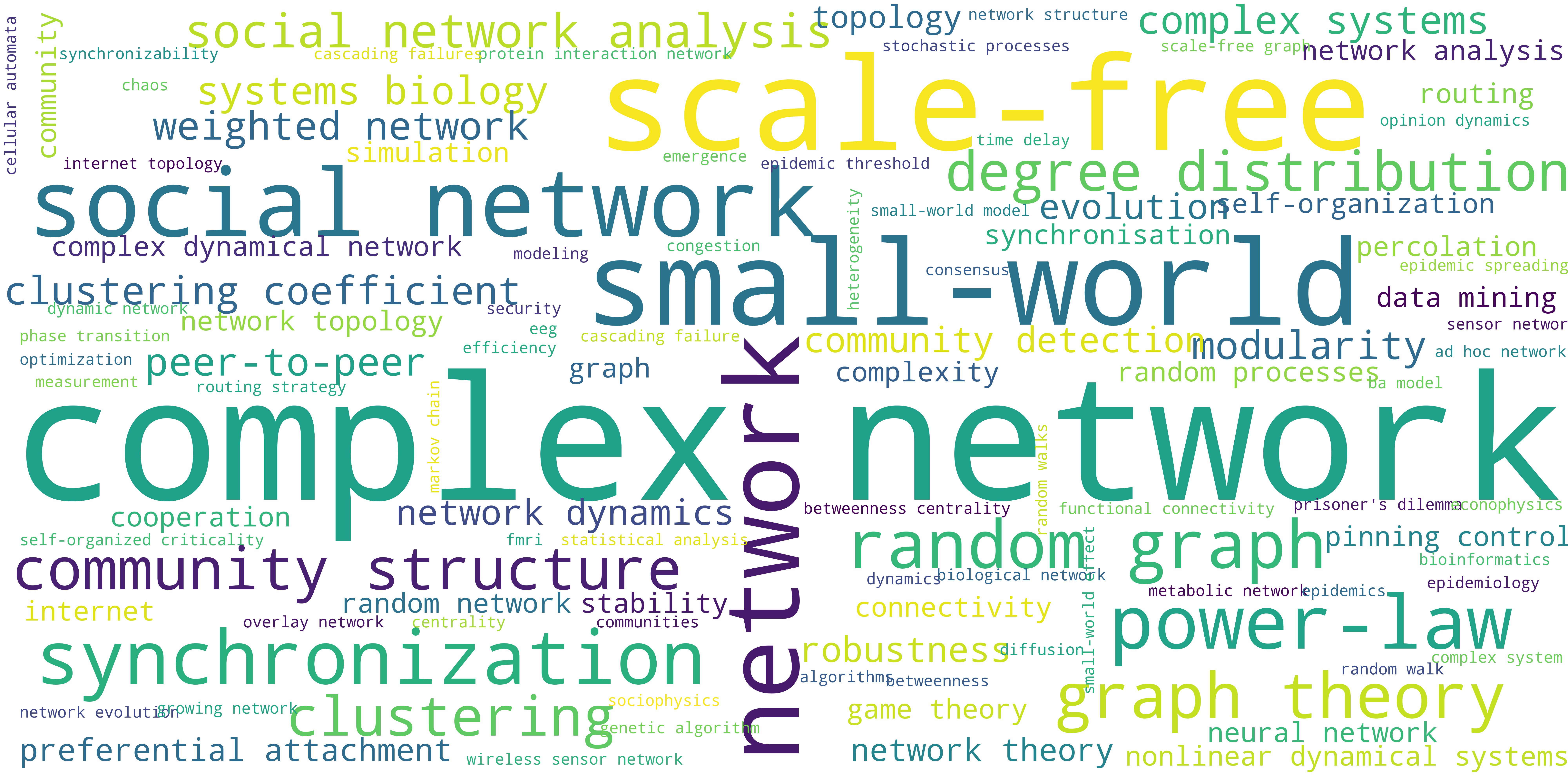}}

    \vspace{0.4cm}
    {\includegraphics[width=0.48\linewidth]{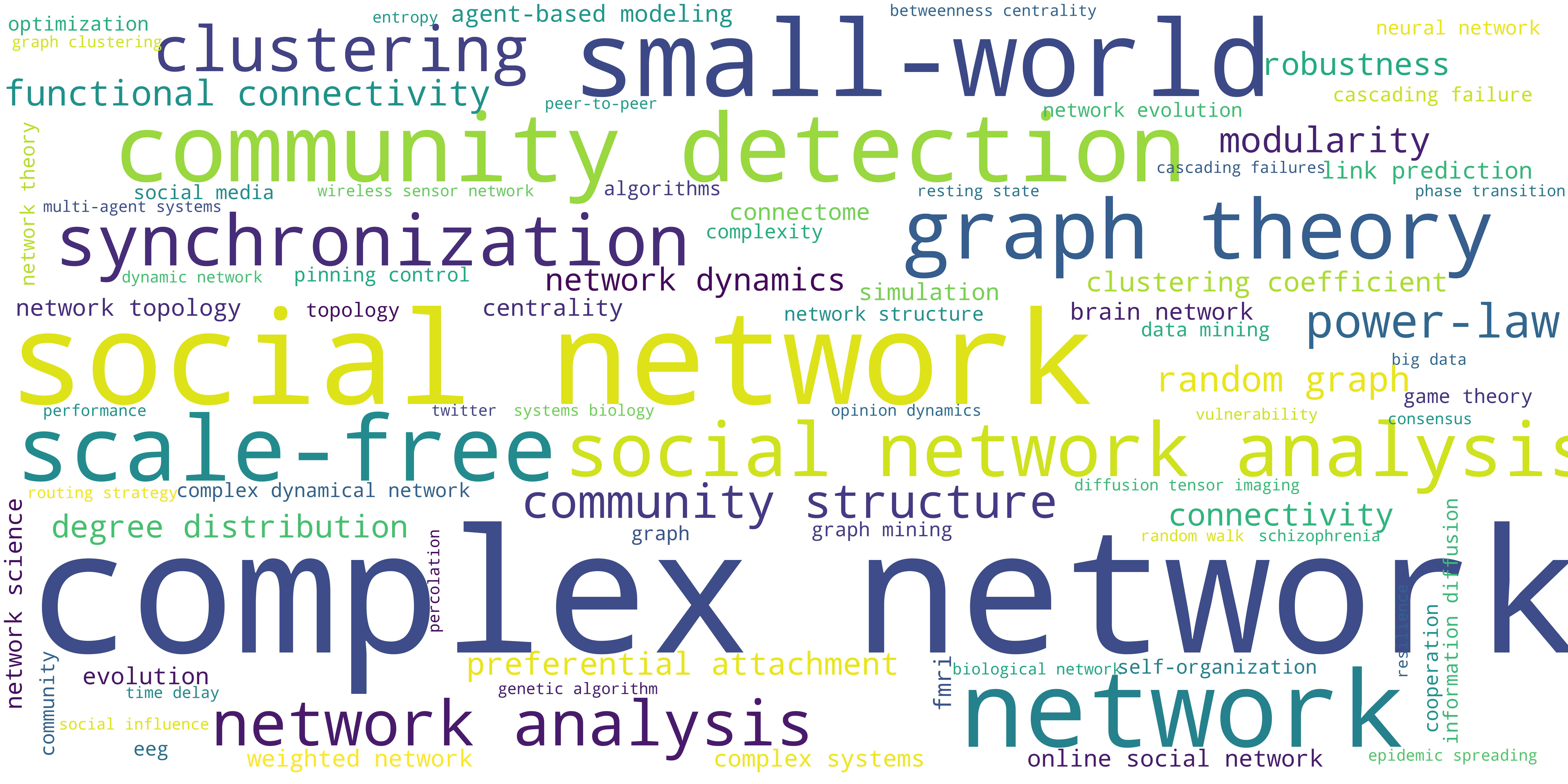}
    \hfill
    \includegraphics[width=0.48\linewidth]{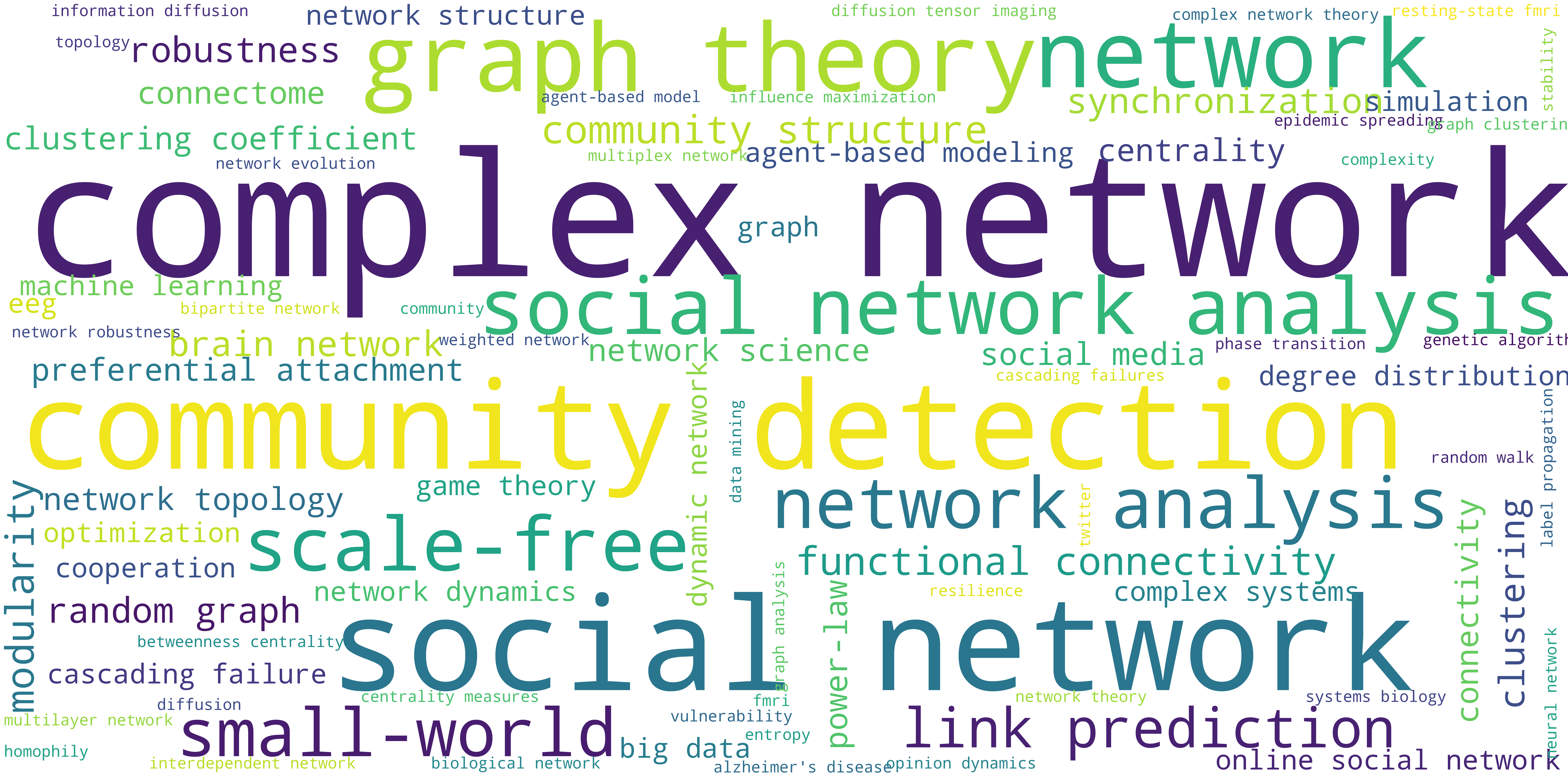}}
    \caption{Word cloud of the most frequent keywords of network science papers before 2006 (top right), between 2006 and 2010 (top left), between 2011 and 2015 (bottom left) and since 2016 (bottom right).}
    \label{wordclouds5}
    \vspace{-12pt}
\end{figure}

Fig.~\ref{wordclouds5} depicts separate word clouds of the most frequently used keywords of network science papers written in the four 5-year-long periods of the last two decades of network science. We can observe that in the first half of the examined period structure related (e.g., scale-free, small-world, topology) and modeling related keywords (e.g., preferential attachment, evolution, growing network, small-world model) dominated the study of complex networks, while in the last decade topics such as community detection, social network analysis, data-driven research (big data, data mining, link prediction, machine learning) have become more popular keywords. In the first decade of the examined period, the most studied real-world networks were Internet, peer-to-peer, and protein interaction networks, however, since 2010 the research has tended to focus on online social networks and brain networks. These word clouds can also be found in the supplementary material~\cite{supp}.

\begin{figure}[h!]
    \centering
    \begin{minipage}{\textwidth}
        \centering
        \includegraphics[width=\textwidth]{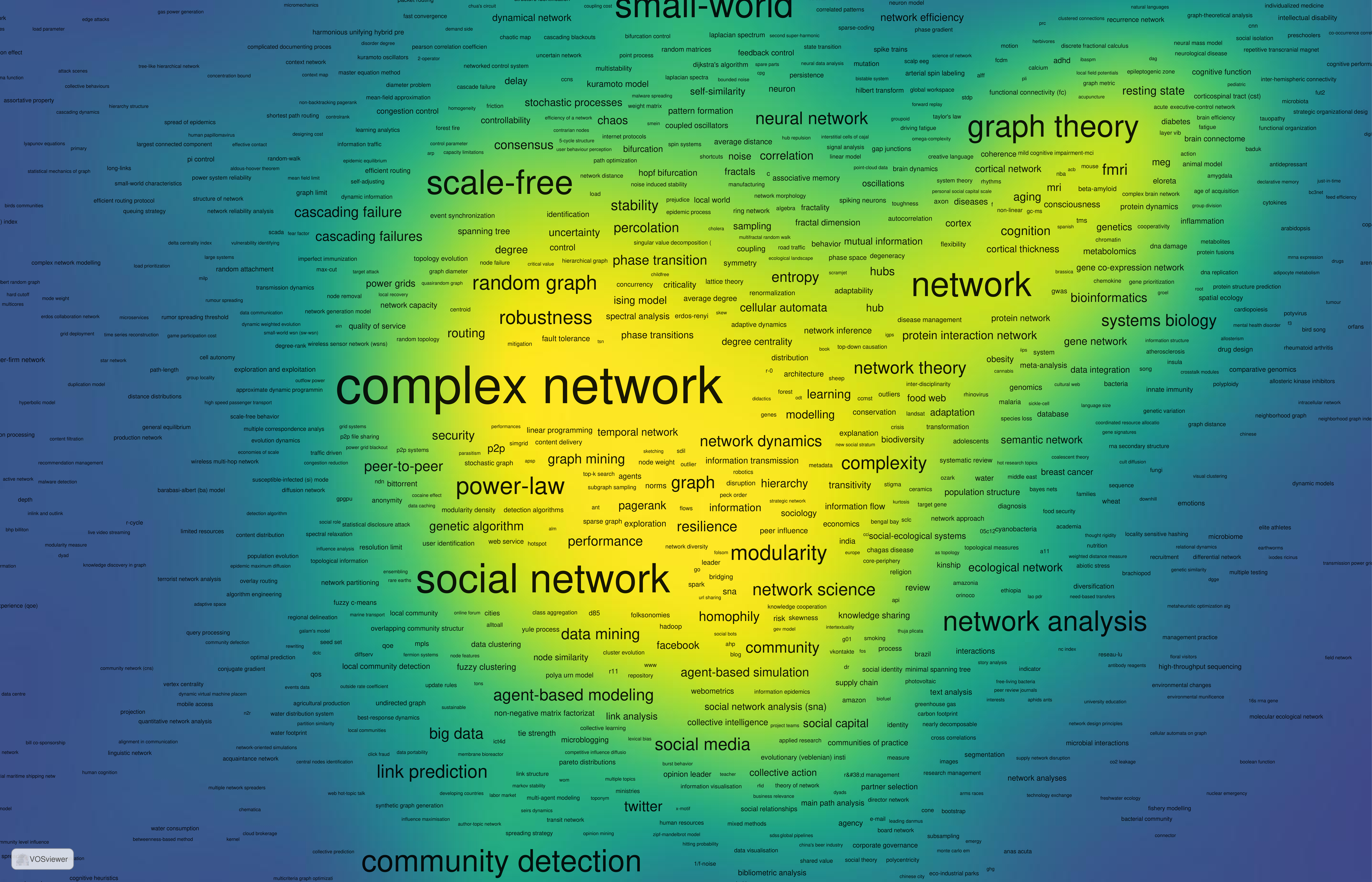}
        \caption{Visualization of two-dimensional knowledge map of keywords. Keywords that have co-occurred more frequently are placed closer to each other on the map. The font size indicates the number and strength of the connections of a keyword. A more intense color implies a larger number of keywords and higher connectivity in the neighborhood of the point. The figure was created with VOSviewer~\cite{van2009software}.}
    \label{fig:knowledge_map}
    \end{minipage}%
    \hfill
    \begin{minipage}{\textwidth}
        \centering
        \includegraphics[width=\textwidth]{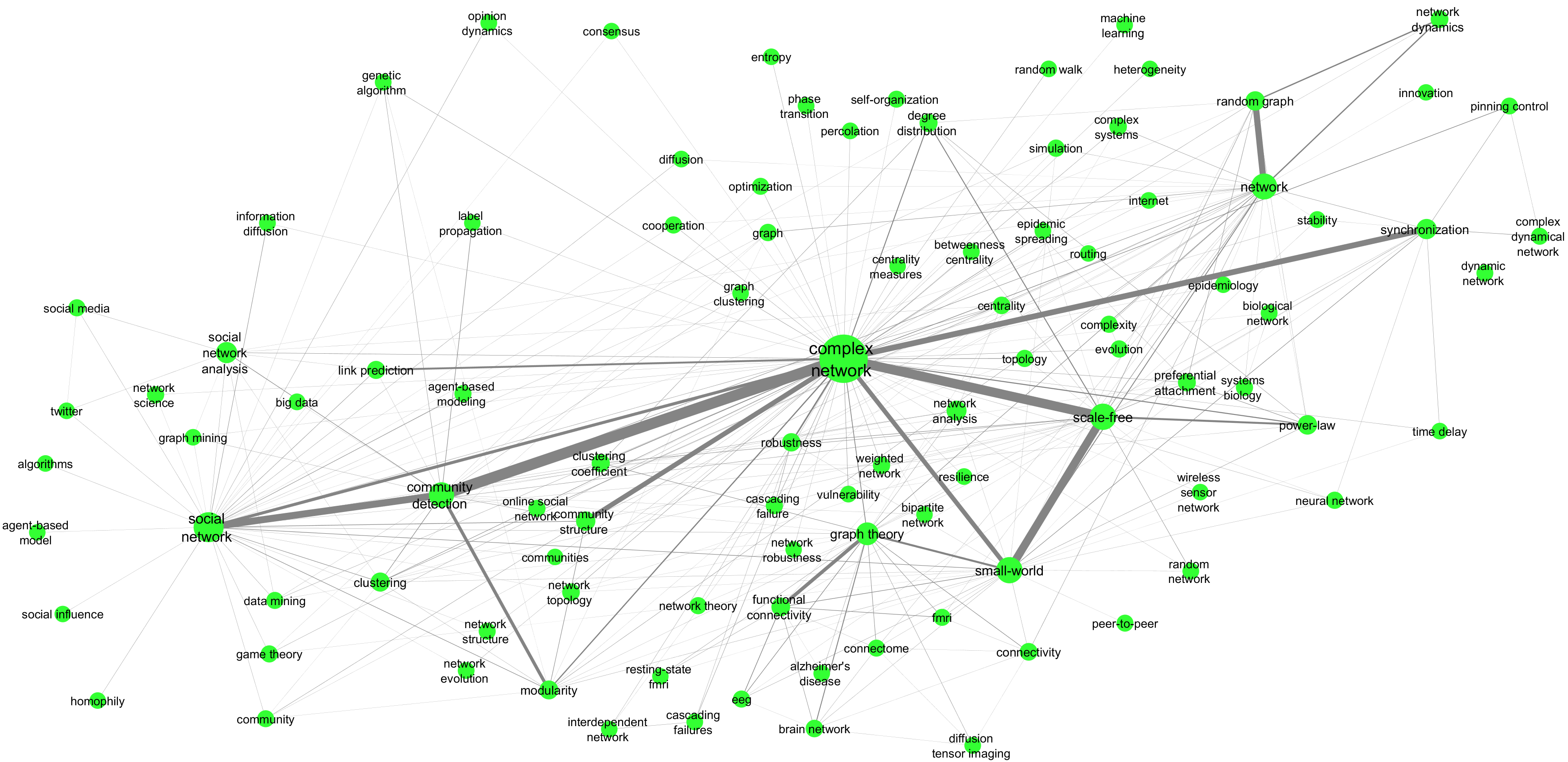}
        \caption{Co-occurrence network of keywords. The size of the node indicates the frequency of keywords in network science papers, the edge width indicates their relative co-occurrence. Only keywords with frequency at least 100 and edges with weight at least 10 are shown in the figure.}
    \label{fig:cooccurence}
    \end{minipage}
\end{figure}

\begin{figure}[h]
    \centering
    \begin{minipage}{0.48\textwidth}
        \centering
        \includegraphics[width=\linewidth]{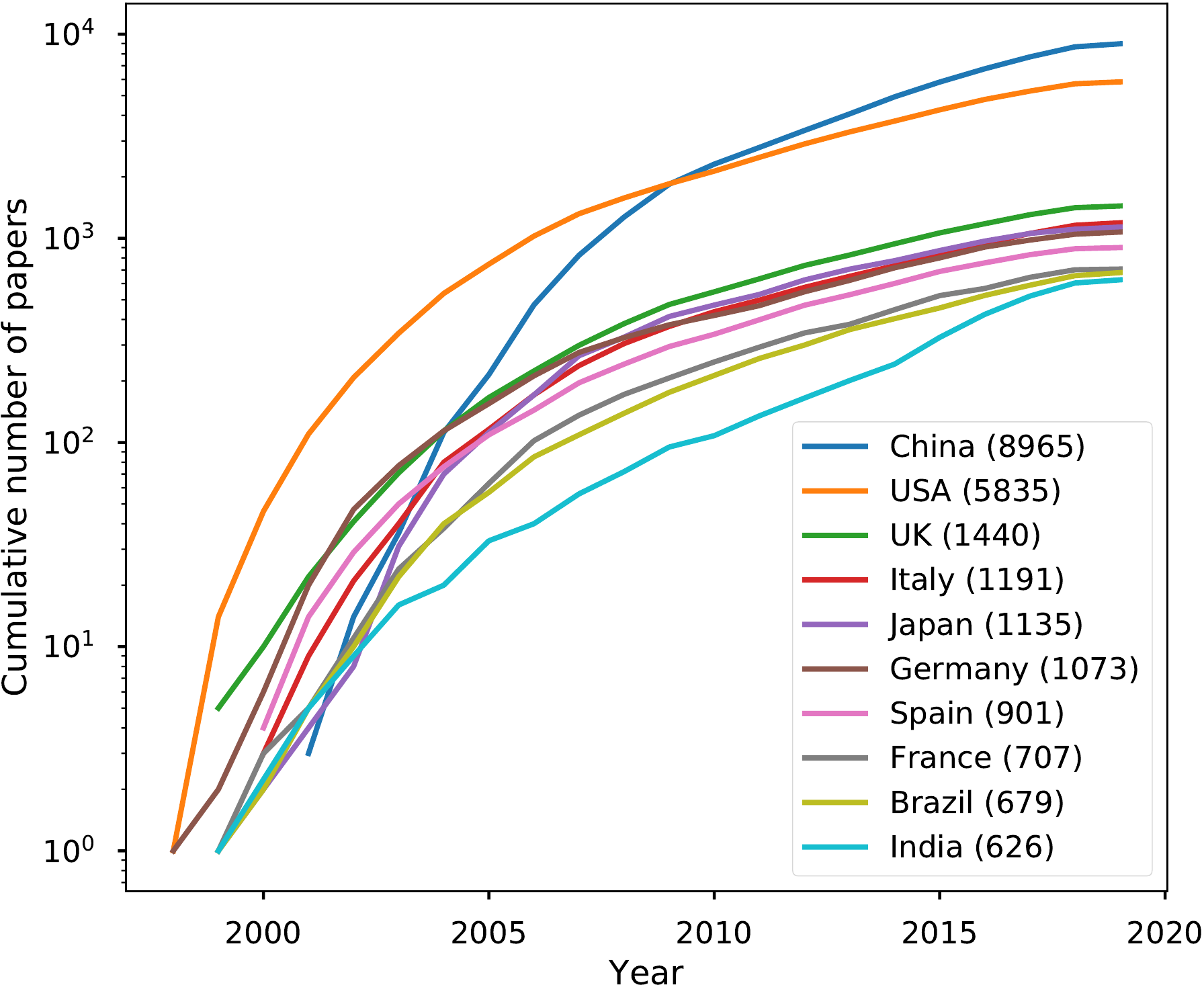}
        \caption{Cumulative number of network science papers on a logarithmic scale by the country of the first author (only the Top 10 countries are shown).}
       % \vspace{-12pt}
        \label{fig:cumsum}
    \end{minipage}%
    \hfill
    \begin{minipage}{0.48\textwidth}
        \centering
        \vspace{-22pt}
        \includegraphics[width=\linewidth]{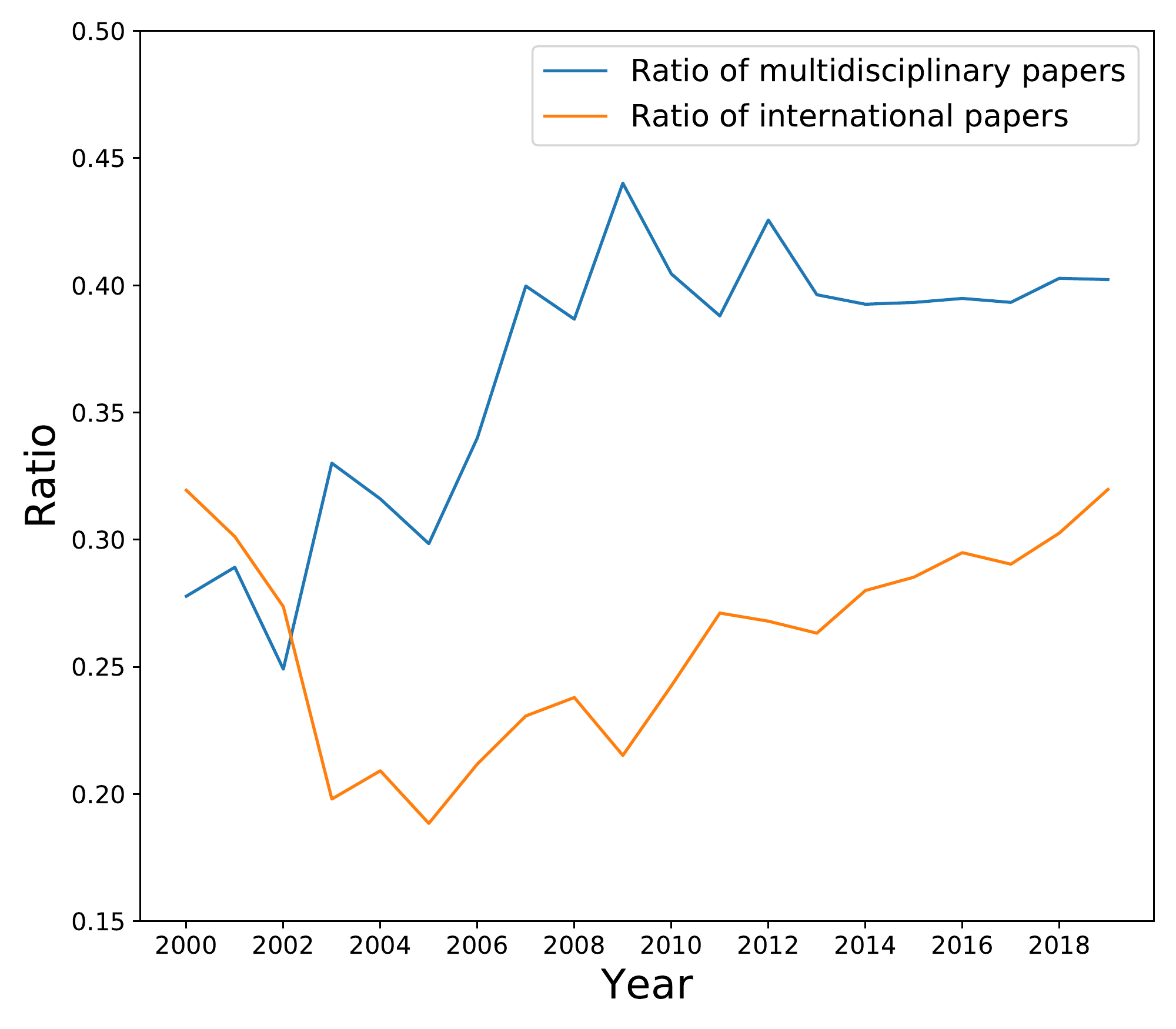}
        \caption{Ratio of multidisciplinary and international papers since 2000.}
        \label{fig:multi_inter}
    \end{minipage}
\end{figure}

To gain a more comprehensive understanding of the keywords of network science papers, we also construct their co-occurrence network. Fig. \ref{fig:knowledge_map} and Fig. \ref{fig:cooccurence} together with  Fig. \ref{wordclouds5} clearly shows that 'complex network' is the most important keyword of the examined set of papers that supports our hypothesis that our definition of network science paper was indeed a good proxy of our purposes. We can observe that the keyword 'random graph' often goes together with 'network' but less often with 'complex network'  that suggests the term 'complex network' is not that widespread among mathematicians. Another observation is that community detection is rather popular in the social network domain, the bottom left side of the figures is dominated by terms associated with social network analysis and community detection. We can also observe that the keywords 'scale-free' and 'small-world' are frequently used together. The complete keyword co-occurrence network and the Figures \ref{fig:knowledge_map} and \ref{fig:cooccurence} can be found in the supplementary material~\cite{supp}.
% \begin{figure}
%     \centering
%     \includegraphics[width=\textwidth]{img/keyword_density.pdf}
%     \caption{Visualization of two-dimensional knowledge map of keywords. Keywords that have co-occurred more frequently are placed closer to each other on the map. The font size indicates the number and strength of the connections of a keyword. A more intense color implies a larger number of keywords and higher connectivity in the neighborhood of the point. The figure was created with VOSviewer~\cite{van2009software}.}
%     \label{fig:knowledge_map}
% \end{figure}
% \begin{figure}
%     \centering
%     \includegraphics[width=\textwidth]{img/keywordgraph.png}
%     \caption{Co-occurrence network of keywords. The size of the node indicates the frequency of keywords in network science papers, the edge width indicates their relative co-occurrence. Only keywords with frequency at least 100 and edges with weight at least 10 are shown in the figure.}
%     \label{fig:cooccurence}
% \end{figure}
Based on the address of the first author, we identify the network science hot-spots and investigate the spatiotemporal changes. Fig.~\ref{fig:cumsum} demonstrates that China and the USA are the two leading countries of network science with a fast increase in Chinese network science papers in the last few years.
% \begin{figure}
% \vspace{-5pt}
%     \centering
%     \includegraphics[width=0.6\linewidth]{img/cum_sum.pdf}
%     \caption{Cumulative number of \textit{network science papers} on a logarithmic scale by the country of the first author (only the Top 10 countries are shown).}
%   % \vspace{-12pt}
%     \label{fig:cumsum}
% \end{figure}

% \begin{figure}
%     \centering
%     \includegraphics[width=0.8\linewidth]{img/multi_inter.pdf}
%     \caption{Ratio of multidisciplinary and international papers}
%     \label{fig:multi_inter}
% \end{figure}

Fig. \ref{fig:multi_inter} illustrates how the ratio of multidisciplinary and international papers varied over the years. We can observe that network science research gets both increasingly international and multidisciplinary. Here we consider a paper international if it has at least two co-authors who do not share an affiliation within the same country. The ratio of international papers in an important indicator, since it was also shown that scientific impact increases if researchers publish in international collaboration \cite{breugelmans2018scientific}. The multidisciplinary nature of papers is defined by the fact that more than one research area is attached to the document in Web of Science.

\section{Analysis of the co-authorship network}
\begin{figure}[h]
    \centering
    \begin{minipage}{0.39\textwidth}
        \centering
         \includegraphics[width=\linewidth]{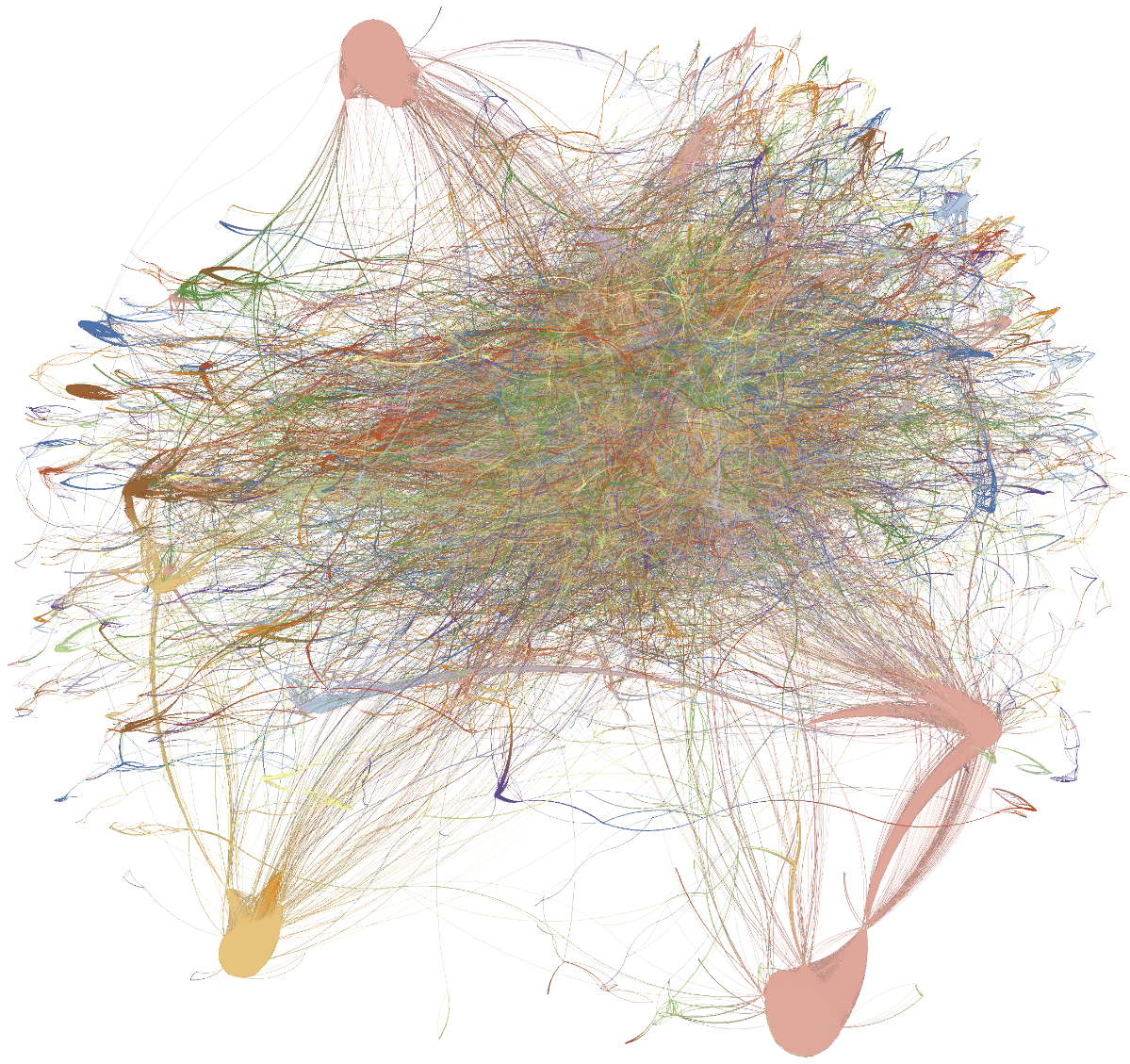}
    \caption{The largest connected component of the co-authorship network of network scientists colored by communities.}
    \label{fig:largest}
    \end{minipage}%
    \hfill
    \begin{minipage}{0.58\textwidth}
        \centering
      \includegraphics[width=\linewidth]{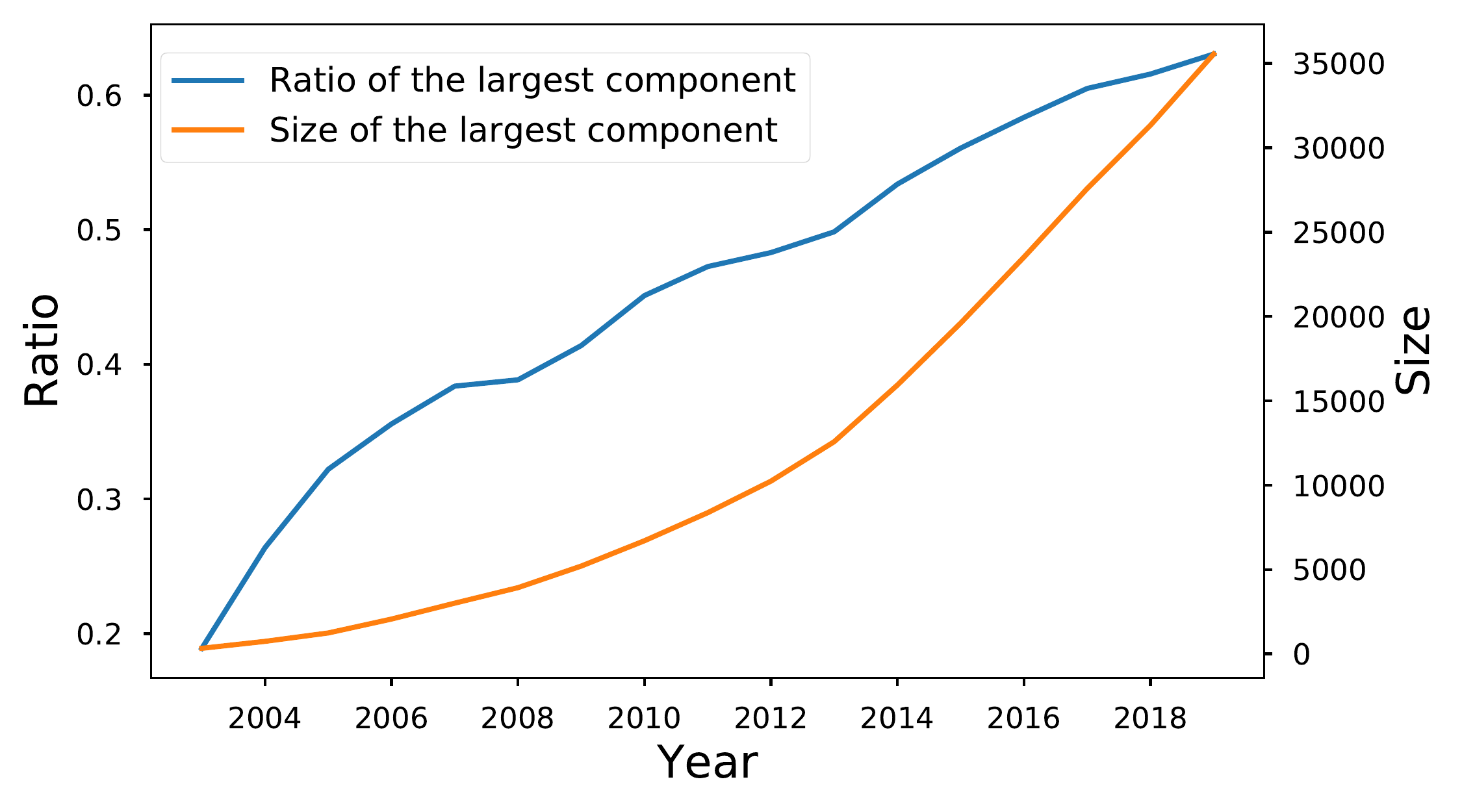}
    % \vspace{-18pt}
    \caption{The absolute and relative size of the largest connected component of the co-authorship network.}
    %\vspace{-10pt}
    \label{fig:size_and_comp}
    \end{minipage}
    \vspace{-12pt}
\end{figure}

The nodes of the co-authorship network of network scientists correspond to the authors who have at least one network science paper (i.e., a paper that cites at least one of the three seminal papers~\cite{barabasi1999emergence,girvan2002community,watts1998collective}), two of them are connected if they co-authored at least one network science paper. The network is simple, undirected, and unweighted meaning that here we ignore the strength of the connection between two scientists, i.e. the number of their joint papers. The network has 56,646 nodes and 357,585 edges with an average degree of 12.63, however, the median degree is just 4. The largest connected component consists of 35,716 nodes and it is depicted in Fig. \ref{fig:largest}.

\begin{figure}[ht]
    \centering
    \begin{minipage}{0.49\textwidth}
    \centering
    \vspace{-14pt}
    \includegraphics[width=\linewidth]{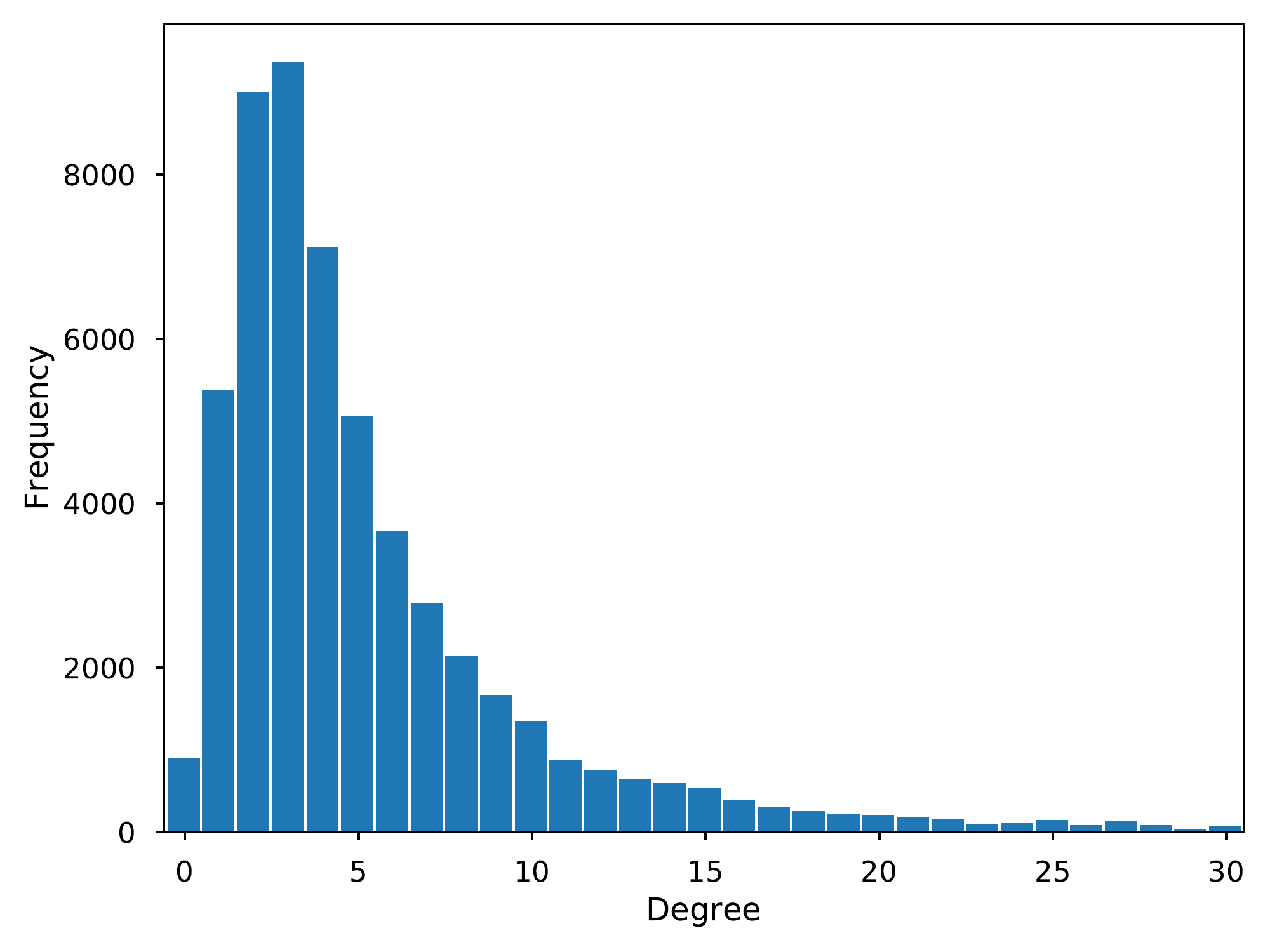}
    % \vspace{-20pt}
    \caption{The degree distribution of the network (truncated at 30).}
    %\vspace{-5pt}
    \label{fig:deg_dist}
    \end{minipage}%
    \hfill
    \begin{minipage}{0.49\textwidth}
       \centering
    \includegraphics[width=\linewidth]{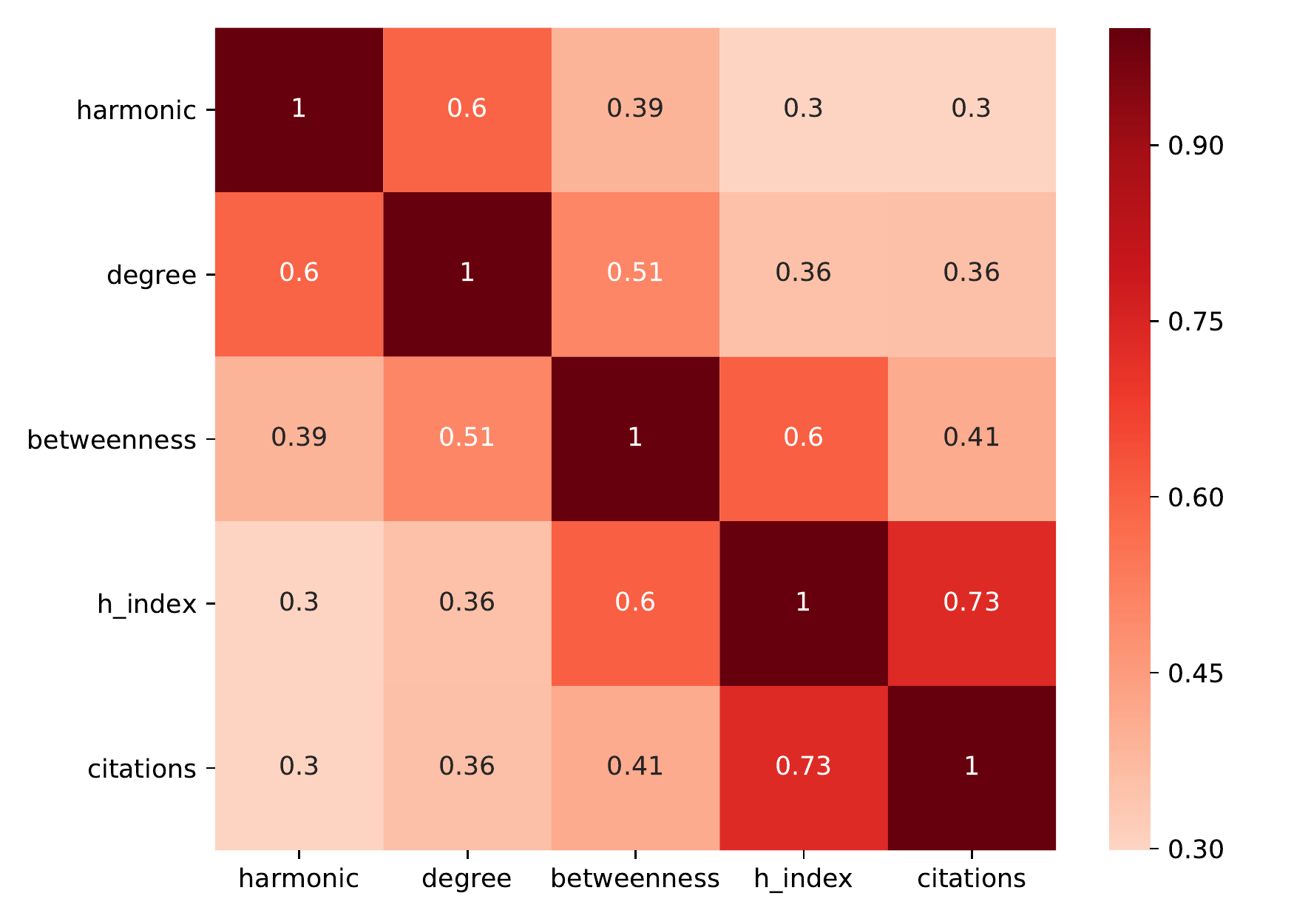}
    \caption{Spearman correlation heatmap between various centrality measures and scientometric indicators.}
    \label{fig:corr_heatmap}
    \end{minipage}
\end{figure}

The degree distribution of the network is illustrated in Fig.~\ref{fig:deg_dist}. There are 897 isolated nodes in the graph (nodes with zero degrees), i.e. scholars who have a single-authored network science paper but have not  co-authored any network science papers. The most typical number of co-authors are between 2 and 4 and the tail of the distribution decays much slower than the number of authors per paper does (c.f. Fig.~\ref{fig:authors_per_paper}) since here the degree reflects all the number of co-authors who do not necessarily author the same paper. The highest degree is 546 corresponding to Roberto Bellotti, a medical physicist, who is also an author of the paper with the highest number of collaborating authors~\cite{lella2018communicability} and another many-authored paper~\cite{choobdar2019assessment}. While our network is unweighted by definition, a possible weight could be assigned to the edges corresponding to the number of joint papers written by the two authors at the endpoints of the edge. Table \ref{table:active_links} shows the most 'active links', i.e. the edges with the highest weights in the edge-weighted version of the co-authorship network.

\begin{table}[h]
\caption{The most active links between authors.}
\label{table:active_links}
\centering
\begin{tabular}{llc}
\multicolumn{2}{c}{\textbf{Authors}} & \multicolumn{1}{l}{\textbf{Number of joint papers}} \\ \hline
Shlomo Havlin & Eugene H. Stanley & 52 \\
Bing-Hong Wang & Tao Zhou & 51 \\
Jihong Guan & Shuigeng Zhou & 50 \\
Zhongzhi Zhang   & Shuigeng Zhou & 48 \\
Jihong Guan  & Zhongzhi Zhang & 40 \\
Zeng-Ru Di   & Ying Fan & 34 \\
Sergey Dorogovtsev  & Jos\'e F.F. Mendes & 32
\end{tabular}
\vspace{-12pt}
\end{table}

The network has a high assortativity coefficient of 0.53 that suggests that nodes tend to be connected to other nodes with similar degrees.  The co-authorship network is highly clustered with a global clustering coefficient of 0.97 and an average local clustering coefficient of 0.8. The fact that the average shortest path length in the largest connected component is 6.6 also supports the small-world nature of co-authorship networks.
% It can be regarded as a small-world network with an average shortest path length of 6.8 in the largest connected component.

% 
\begin{table}[h]
\caption{The top 12 authors with the highest betweenness centrality. Their ranks with respect to each metric are shown in brackets.}
\label{table:centralities}
\centering
\begin{minipage}{\textwidth}
% \begin{savenotes}
\renewcommand{\thefootnote}{\arabic{footnote}}
\begin{tabular}{llllll}
\multicolumn{1}{c}{\multirow{2}{*}{\textbf{Name}}} & \multicolumn{3}{c}{\textbf{Centralities}} & \multirow{2}{*}{\textbf{\begin{tabular}[c]{@{}c@{}}Number of\\ citations\end{tabular}}} & \multirow{2}{*}{\textbf{$h$-index}} \\
\multicolumn{1}{c}{} & \multicolumn{1}{l}{\textbf{Betweenness}} & \textbf{Harmonic} & \textbf{Degree}  \\ \hline
J\"urgen Kurths & \textbf{0.025 (1)} & \textbf{0.169 (1)} & 216 (1,017) & 9,249 (30) & 96 (2)\\
H. Eugene Stanley & 0.024 (2) & 0.168 (2) & 220 (1,013) & 10,479 (18) & 57 (7)\\
Guanrong Chen & 0.019 (3) & 0.165 (4) & 215 (1,018) & 12,859  (15) & 27 (30)\\
Albert-L\'aszl\'o Barab\'asi & 0.017 (4) & 0.160 (12) & 202 (1,023) & \textbf{73,937 (2)} & 83 (3)\\
Yong He & 0.014 (5) & 0.163 (6) & \textbf{242 (1,012)} & 9,104 (32) & 61 (6) \\
Zhen Wang & 0.014 (6) & 0.163 (5) & 155 (1,117) & 3,306 (365) & 39 (16)\\
Santo Fortunato & 0.013 (7) & 0.160 (13) & 208 (1,021) & 13,923 (12) & 40 (14) \\
Shlomo Havlin & 0.013 (8) & 0.163 (7) & 165 (1,042) & 13,377 (13) & \textbf{110 (1)}\\
Tao Zhou & 0.013 (9) & 0.167 (3) & 220 (1,013) & 9,911 (20) & 40 (14)\\
Edward T. Bullmore & 0.012 (10) & 0.151 (49) & 210 (1,020) & 17,915 (7) & 50 (10)  \\
% Wei Wang\footnote[1]{Sichuan University} & 0.012 (11) & 0.161 (10) & 145 (1,178) & 467 (1511) & 14 (188)\\
Wei Wang\footnotemark[1] & 0.012 (11) & 0.161 (10) & 145 (1,178) & 467 (1511) & 14 (188)\\
Stefano Boccaletti & 0.112 (12) & 0.162 (9) & 130 (1,179) & 9,609 (21) & 22 (58)
%\vspace{-15pt}
\end{tabular}
\footnotetext{\hspace{-12pt} \textsuperscript{1}Sichuan University}
% \end{savenotes}
\end{minipage}
\vspace{-12pt}
\end{table}

To identify the most central authors of the network science community as seen through the co-authorship network, we calculate centrality measures such as betweenness, harmonic and degree centralities of the nodes. The most central authors are shown in Table~\ref{table:centralities}. We also compare the centrality measures of the authors with the citation count of their network science papers and with their $h$-indices (restricted only to their network science papers). Common characteristics of the most central authors that they are famous, well-established researchers, moreover, they are typically active in more research areas forming bridges between subdisciplines. The highest betweenness and harmonic centralities correspond to J\"urgen Kurths, German physicist and mathematician whose research is mainly concerned with nonlinear physics and complex systems sciences. As we mentioned before, the highest degree corresponds to Roberto Bellotti, a medical physicist. Mark Newman English-American physicist has the highest number of citations on his network science papers (77,418), while Shlomo Havlin, Israeli physicist is ranked first with respect to $h$-index.

\begin{figure}[h]
% \vspace{-6pt}
    \centering
    \includegraphics[width=0.49\linewidth]{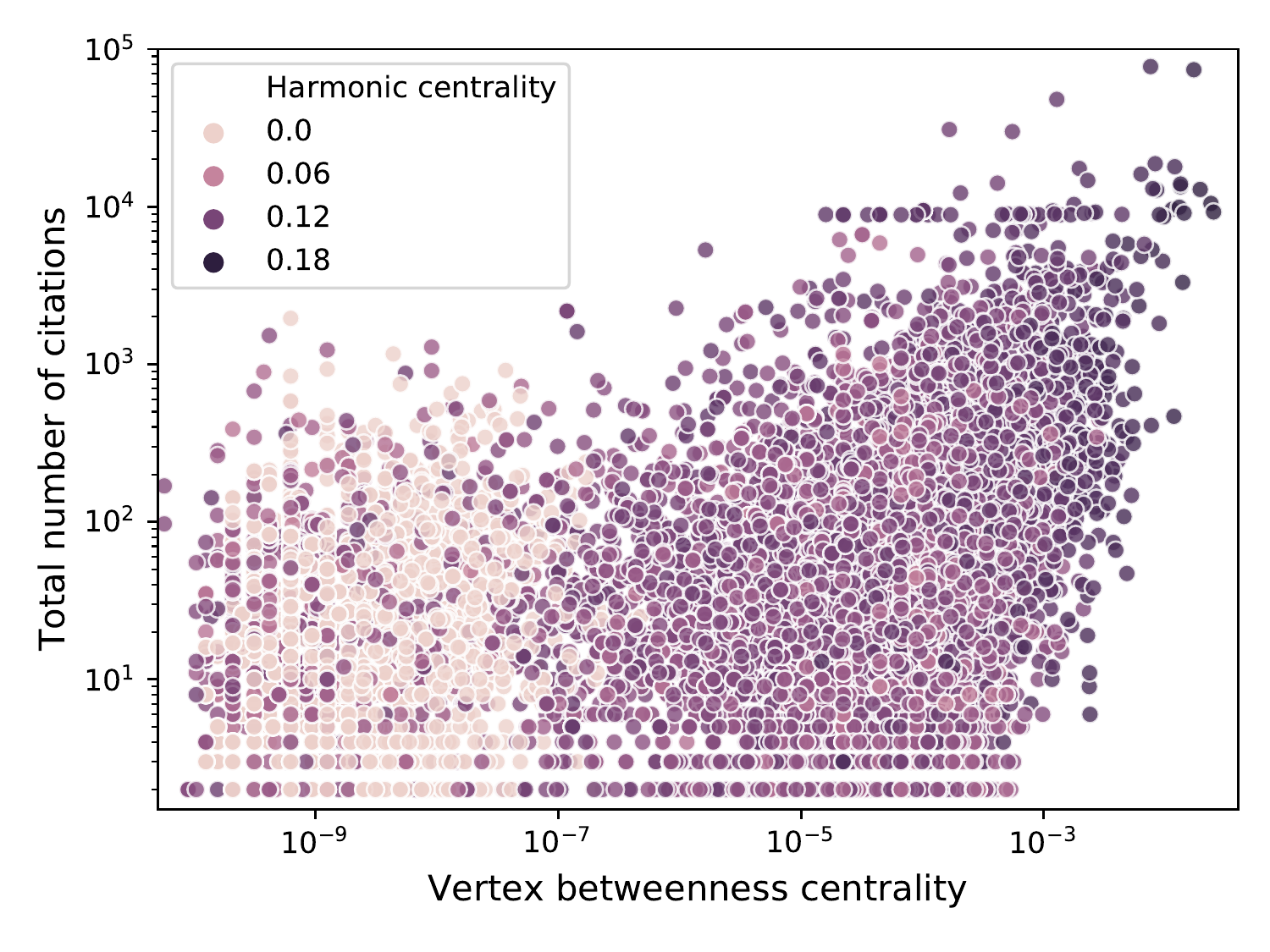}
    \includegraphics[width=0.49\linewidth]{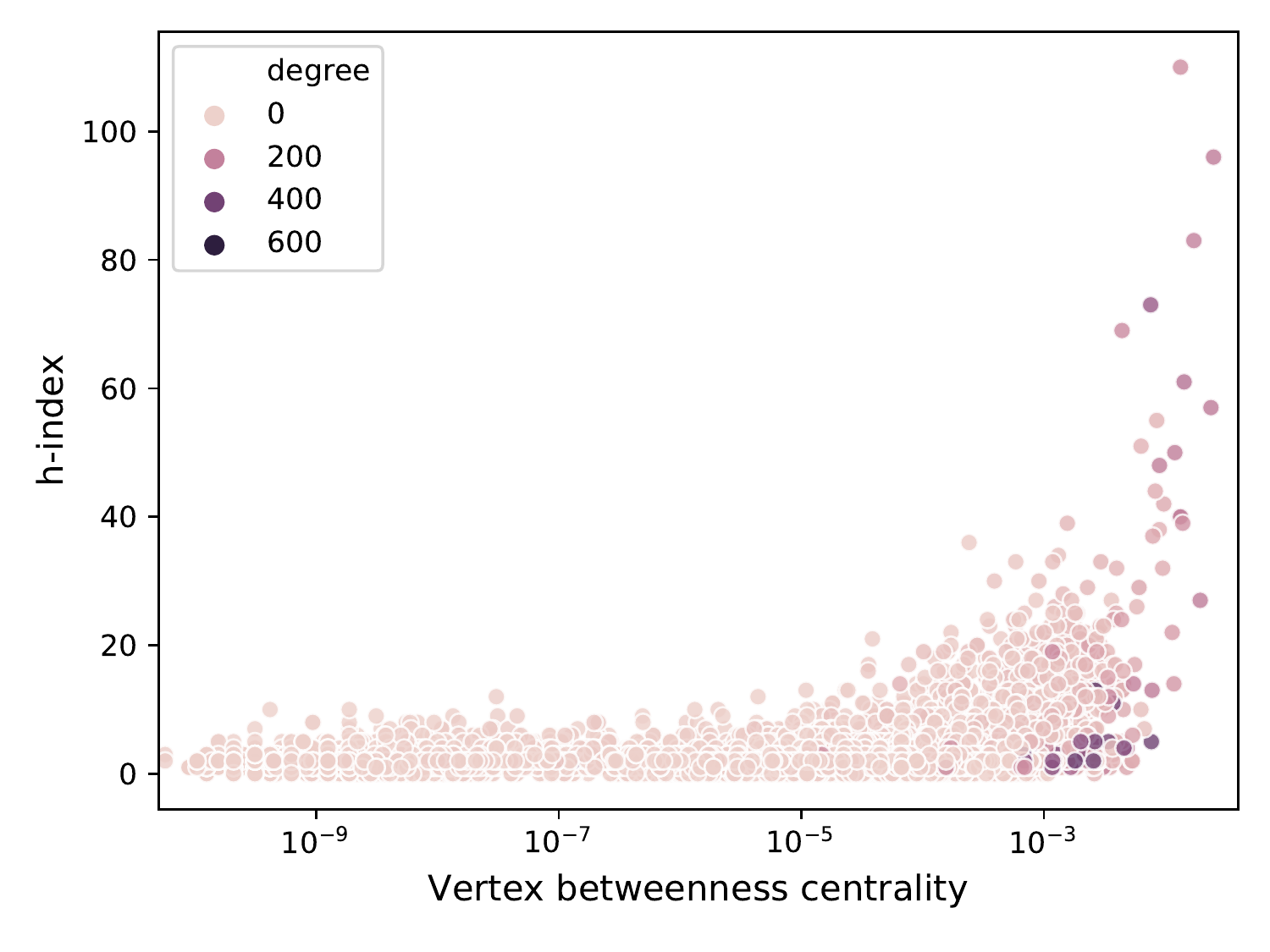}
    % \vspace{-5pt}
    \caption{Relationship between centrality measures of network scientists and the scientometric indicators of their network science papers.}
    \vspace{-12pt}
    \label{fig:centralities}
\end{figure}

Fig.~\ref{fig:centralities} illustrates the relationship between centrality measures of network scientists and the scientometric indicators of their network science papers. On the left it shows the number of citations against the vertex betweenness centrality, colored by the harmonic centrality; on the right one can see the $h$-index against the vertex betweenness centrality, colored by the degree.  We can conclude that there is a positive correlation between the authors' central role in the co-authorship network and their scientometric indicators.  Fig. \ref{fig:corr_heatmap}  shows the Spearman's rank correlation heatmap of the aforementioned measures indicating positive correlations, with the highest positive correlation between citation count and $h$-index. Considering centrality measures against scientometric indicators, betweenness centrality and $h$-index has the highest correlation.

Network scientists have become more connected as time has gone by, as it is illustrated in Fig. \ref{fig:size_and_comp}, since not only the size of the largest component has increased over the years but also the ratio of the size of the giant component to the size of the entire network, indicating the emergence of a diverse but not divided network science community. The giant component consists of 35,716 nodes that is 63\% of the entire network size and it is illustrated in Fig.~\ref{fig:largest}.

\begin{table}[h]
% \vspace{-1cm}
\caption{Composition of the largest communities.}
\vspace{-3pt}
\centering
\label{tab:comm}
\begin{tabular}{lllllll}
\textbf{Size} & \multicolumn{3}{c}{\textbf{Research area}} & \multicolumn{3}{c}{\textbf{Country/Region}} \\ \hline
15,693 & \textbf{PHY 31\%} & \textbf{CS  30\%} & NN 10\% & \textbf{CHN 55\%} & EU 14\% & USA 12\% \\
1,066 & \textbf{CS 31\%} & PHY 15\% & BMB 12\% & CHN 31\% & USA 26\% & EU 17\% \\
812 & \textbf{NN 72\%} & CS 5\% & PSY 4\% & \textbf{USA 90\%} & CAN 3\% & CHN 2\% \\
759 & \textbf{CS 30\%} & PHY 19\% & BMB 16\% & EU 36\% & USA 29\% & ISR 8\% \\
756 & NN 23\% & CS 17\% & PHY 16\% & EU 34\% & JPN 17\% & KOR 15\% \\
711 & \textbf{CS 30\%} & MAT 10\% & LSB 7\%  & USA 35\% & EU 29\% & CHN 9\% \\
633 & CS 19\% & PHY 18\%  & BMB 8\% & CHN 27\% & EU 23\% & IND 13\% \\
563 & \textbf{ESE 44\%} & CS 11\% & LSB 8\% & \textbf{EU 53\%} & BRA 12\% & USA 12\% \\
559 & \textbf{CS 42\%} & ENG 14\% & PHY 12\% & USA 30\% & EU 20\% & IRN 11\% \\ 
555 & BMB 35\% & CS 23\%  & MCB 8\% & USA 30\% & EU 24\% & CHN 21\% 
%523 & SCT 51\%  & CS 16\%  & BMB 11\% & USA 28\% & IND 16\% & EU 12\% \\
%512 & CS 28\% & NN 13\% & BMB 10\% & \textbf{EU 44 \%} & USA 23\% & IND 6\% 
\\\hline
\multicolumn{4}{l}{\scriptsize \textbf{ACS}: Automations \& Control Systems} & \multicolumn{3}{l}{\scriptsize\textbf{BE}: Business \& Economics} \\
\multicolumn{4}{l}{\scriptsize \textbf{BMB}: Biochemistry \& Molecular Biology} & \multicolumn{3}{l}{\scriptsize\textbf{CS}: Computer Science} \\
\multicolumn{4}{l}{\scriptsize\textbf{ESE}: Environmental Sciences \& Ecology} & \multicolumn{3}{l}{\scriptsize\textbf{GH}: Genetics \& Heredity} \\
\multicolumn{4}{l}{\scriptsize\textbf{LSB}: Life Sciences \& Biomedicine} & \multicolumn{3}{l}{\scriptsize\textbf{MAT}: Mathematics} \\
\multicolumn{4}{l}{\scriptsize\textbf{MCB}: Mathematical \& Computational Biology} & \multicolumn{3}{l}{\scriptsize\textbf{NN}: Neurosciences \& Neurology} \\
\multicolumn{4}{l}{\scriptsize\textbf{PHY}: Physics} & \multicolumn{3}{l}{\scriptsize\textbf{PSY}: Psychiatry}  \\
\multicolumn{4}{l}{\scriptsize\textbf{SCT}: Science \& Technology}& \multicolumn{3}{l}{\scriptsize\textbf{TEL}: Telecommunication}\\
\multicolumn{7}{l}{\scriptsize The country abbreviations are the officially assigned ISO alpha-3 codes \cite{iso_country_codes}}
\end{tabular}
\vspace{-12pt}
\end{table}

Using Clauset-Newman-Moore greedy modularity maximization community detection algorithm \cite{clauset2004finding}, we identify the dense subgraphs of the network. To retrieve some important discipline and location-related characteristics of the largest communities, we assigned a research area and a country for each author as the majority of the research areas corresponding to their papers and the most frequent country of their affiliations respectively.  The compositions of the ten largest communities are shown in Table~\ref{tab:comm}. The largest community consists of 15,693 authors dominated by Chinese physicists and computer scientists. We can observe that the smaller the communities are, the more homogeneous they are. For example, the vast majority of the third-largest community are  North American neuroscientists, moreover, there is a community with 53\% EU scientists and 44\% environmental scientists.

\begin{figure}[h]
    \centering
    \begin{minipage}{0.54\textwidth}
        \centering
         \includegraphics[width=\linewidth]{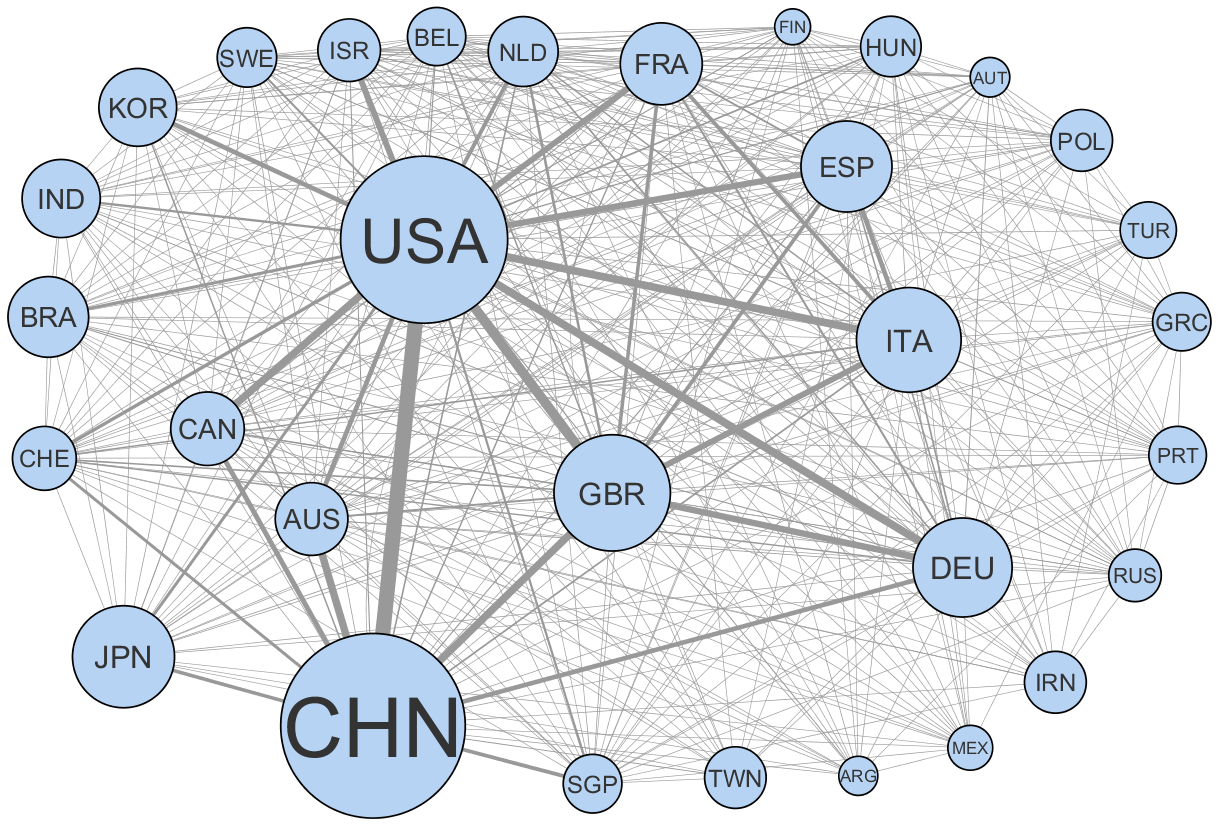}
    % \vspace{-12pt}
    \caption{Network of international collaborations. The size of the node corresponds to the number of network science papers authored by at least one scientist from the corresponding country, the edge width indicates the number of papers written in the collaboration of authors from the corresponding countries. Only countries with at least 100 network science papers are shown in the figure.}
    \label{country_graph}
    \end{minipage}%
    \hfill
    \begin{minipage}{0.42\textwidth}
        \centering
      \includegraphics[width=\linewidth]{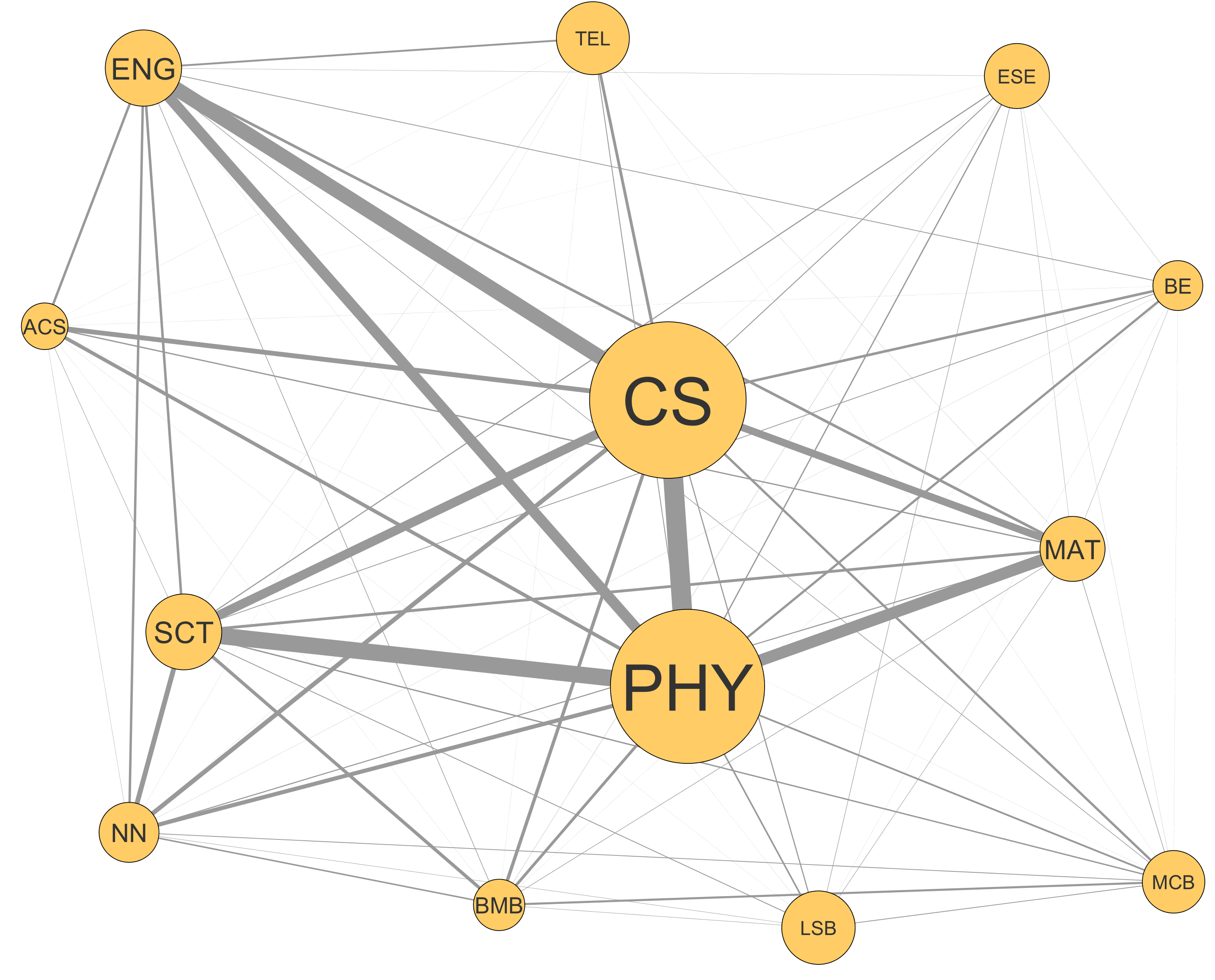}
  \captionof{figure}{Network of multidisciplinary collaborations. Only the research areas formed by at least 500 network scientists are shown in the figure. The full names of the research areas can be found in Table \ref{tab:comm}.}
  \label{fig:ra_graph}
    \end{minipage}
\end{figure}

Network scientists come from 118 different countries which shows the international significance of network science. To illustrate the typical patterns of international collaborations, we created an edge-weighted network of countries where edge weights correspond to the number of network science papers that were written in the collaboration of at least one author from both countries (see Fig.~\ref{country_graph}). We can observe that while China has the highest number of network science papers (see also Fig.~\ref{fig:cumsum}), US scientists wrote the most articles in international collaboration. It is also apparent that EU countries collaborate with each other a lot.

Similarly to the network of international collaborations, we also created a network of multidisciplinary collaborations illustrating the importance of multidisciplinary research in network science. Fig. \ref{fig:ra_graph} shows an edge-weighted network of research areas where the edge weights correspond to the number of network science papers that were written in the collaboration of authors whose main research areas are the ones at the endpoints of the edge. The main research area of the authors is not given in the Web of Science, so for each author, we assigned the most frequent research area associated with their papers. We can observe that computer scientists and physicists dominate network science. It is also clear that the collaboration of physicists and network scientists made huge progress in network science. We can conclude that --~as far as network science papers are concerned~-- mathematicians collaborate the most with physicists, while engineers collaborate more with computer scientists. It is not surprising that telecommunication experts usually collaborate with engineers and computer scientists, while mathematical \& computational biologists work a lot with biochemists \& molecular biologists and computer scientists on network science papers.

% \begin{figure}
% \centering
% \begin{minipage}{.5\textwidth}
%   \centering
%   \includegraphics[width=.8\linewidth]{img/cnt_graph.pdf}
%   \captionof{figure}{A figure}
%   \label{fig:test1}
% \end{minipage}%
% \begin{minipage}{.5\textwidth}
%   \centering
%   \includegraphics[width=.7\linewidth]{img/ra_areas_graph.pdf}
%   \captionof{figure}{Another figure}
%   \label{fig:test2}
% \end{minipage}
% \end{figure}

\section{Conclusion}

Two decades ago a new multidisciplinary scientific field was born: network science. In this paper, we paid tribute to the network science community by investigating the past 20 years of complex network research as seen through the co-authorship network of network scientists. We studied 31,763 network science papers by extracting the distributions of research areas, journals, and keywords. We identified the most important publication venues and topics in network science and shed light on how they changed over time, we also explored the co-occurrence network of the keywords.  Moreover, we constructed and extensively analyzed the co-authorship network of 56,646 network scientists, for example investigating its topological properties, namely its community structure, degree, and centrality distributions. We identified the most central authors of network science as seen through the co-authorship network. We also studied the spatiotemporal changes to provide insights on collaboration patterns. We can conclude that both international and interdisciplinary collaborations are on the increase and the network science community is getting more connected. Furthermore, we compared the centrality measures of authors with well-known scientometric indicators (e.g. citation count and $h$-index) and found a high correlation.

The present study also has its own limitations. Most importantly, our definitions of a network science paper and a network scientist are quite arbitrary but we believe our chosen notions are good proxies for the purpose of this study that is also supported by the distribution of the keywords of the examined papers. Additionally, the data set itself is not consistent due to different naming conventions that we aimed to resolve, furthermore, we cannot distinguish between different scientists with the same name. However, the error introduced by these problems is negligible. 

After investigating the publication and collaboration patterns of network science and observing an increasing impact of complex networks, we are convinced that the next 20 years will produce at least as many fruitful scientific collaborations and outstanding discoveries in network science as the last two decades.
\afterpage{\clearpage}
\section*{Acknowledgment}
%\vspace{-0.05cm}

We thank the anonymous reviewers of the Lecture Notes in Social Networks for their observations and comments on an earlier version of this paper. 
The research reported in this paper was supported by the BME-Artificial Intelligence FIKP grant of EMMI (BME FIKP-MI/SC). The publication is also supported by the EFOP-3.6.2-16-2017-00015 project entitled 'Deepening the activities of HU-MATHS-IN, the Hungarian Service Network for Mathematics in Industry and Innovations'. The research of Roland Molontay was partially supported by the NKFIH K123782 research grant.

% the environments 'definition', 'lemma', 'proposition', 'corollary',
% 'remark', and 'example' are defined in the LLNCS documentclass as well.
%
% For citations of references, we prefer the use of square brackets
% and consecutive numbers. Citations using labels or the author/year
% convention are also acceptable. The following bibliography provides
% a sample reference list with entries for journal
% articles~\cite{ref_article1}, an LNCS chapter~\cite{ref_lncs1}, a
% book~\cite{ref_book1}, proceedings without editors~\cite{ref_proc1},
% and a homepage~\cite{ref_url1}. Multiple citations are grouped
% \cite{ref_article1,ref_lncs1,ref_book1},
% \cite{ref_article1,ref_book1,ref_proc1,ref_url1}.
%
% ---- Bibliography ----
%
% BibTeX users should specify bibliography style 'splncs04'.
% References will then be sorted and formatted in the correct style.
%
\bibliographystyle{splncs04}
\bibliography{mybib}

\begin{thebibliography}{10}
\providecommand{\url}[1]{\texttt{#1}}
\providecommand{\urlprefix}{URL }
\providecommand{\doi}[1]{https://doi.org/#1}

\bibitem{barabas2019co}
Barab\'as, B., F\"ul\"op, O., Molontay, R.: The co-authorship network and
  scientific impact of {L}ászló {L}ovász. Journal of Combinatorial
  Mathematics and Combinatorial Computing  \textbf{108},  187--192 (2019)

\bibitem{barabas2017impact}
Barab{\'a}s, B., F\"ul\"op, O., Molontay, R., P{\'a}lyi, G.: Impact of the
  discovery of fluorous biphasic systems on chemistry: {A} statistical and
  network analysis. ACS Sustainable Chemistry \& Engineering  \textbf{5}(9),
  8108--8118 (2017)

\bibitem{barabasi2019twenty}
Barab\'asi, A.: Twenty years of network science: From structure to control.
  Bulletin of the American Physical Society  (2019)

\bibitem{barabasi2003linked}
Barab{\'a}si, A.L.: Linked: {T}he new science of networks (2003)

\bibitem{barabasi1999emergence}
Barab{\'a}si, A.L., Albert, R.: Emergence of scaling in random networks.
  Science  \textbf{286}(5439),  509--512 (1999)

\bibitem{barabasi2002evolution}
Barab{\'a}si, A.L., Jeong, H., N{\'e}da, Z., Ravasz, E., Schubert, A., Vicsek,
  T.: Evolution of the social network of scientific collaborations. Physica A:
  Statistical Mechanics and its Applications  \textbf{311}(3-4),  590--614
  (2002)

\bibitem{barabasi2016network}
Barab{\'a}si, A.L., et~al.: Network science. Cambridge Univ. Press (2016)

\bibitem{breugelmans2018scientific}
Breugelmans, J.G., Roberge, G., Tippett, C., Durning, M., Struck, D.B.,
  Makanga, M.M.: Scientific impact increases when researchers publish in open
  access and international collaboration: {A} bibliometric analysis on
  poverty-related disease papers. PloS one  \textbf{13}(9),  e0203156 (2018)

\bibitem{choobdar2019assessment}
Choobdar, S., Ahsen, M.E., Crawford, J., Tomasoni, M., Fang, T., Lamparter, D.,
  Lin, J., Hescott, B., Hu, X., Mercer, J., et~al.: Assessment of network
  module identification across complex diseases. Nature methods
  \textbf{16}(9),  843--852 (2019)

\bibitem{clauset2004finding}
Clauset, A., Newman, M.E., Moore, C.: Finding community structure in very large
  networks. Physical Review E  \textbf{70}(6),  066111 (2004)

\bibitem{national2005Network}
Council, N.R., et~al.: Network science committee on network science for future
  army applications (2005)

\bibitem{fortunato2018science}
Fortunato, S., Bergstrom, C.T., B{\"o}rner, K., Evans, J.A., Helbing, D.,
  Milojevi{\'c}, S., Petersen, A.M., Radicchi, F., Sinatra, R., Uzzi, B.,
  et~al.: Science of science. Science  \textbf{359}(6379),  eaao0185 (2018)

\bibitem{girvan2002community}
Girvan, M., Newman, M.E.: Community structure in social and biological
  networks. Proceedings of the National Academy of Sciences  \textbf{99}(12),
  7821--7826 (2002)

\bibitem{iso_country_codes}
{International Organization for Standardization}: Officially assigned ISO
  3166-1 alpha-3 codes (2020), \url{https://www.iso.org/obp/ui/}

\bibitem{kocarev2010network}
Kocarev, L., In, V.: Network science: {A} new paradigm shift. IEEE Network
  \textbf{24}(6) (2010)

\bibitem{kumar2015co}
Kumar, S.: Co-authorship networks: a review of the literature. Aslib Journal of
  Information Management  \textbf{67}(1),  55--73 (2015)

\bibitem{lella2018communicability}
Lella, E., Amoroso, N., Lombardi, A., Maggipinto, T., Tangaro, S., Bellotti,
  R., Initiative, A.D.N.: Communicability disruption in {A}lzheimer’s disease
  connectivity networks. Journal of Complex Networks  \textbf{7}(1),  83--100
  (2018)

\bibitem{leonidou2010five}
Leonidou, L.C., Katsikeas, C.S., Coudounaris, D.N.: Five decades of business
  research into exporting: A bibliographic analysis. Journal of International
  Management  \textbf{16}(1),  78--91 (2010)

\bibitem{li2016evolutionary}
Li, H., An, H., Wang, Y., Huang, J., Gao, X.: Evolutionary features of academic
  articles co-keyword network and keywords co-occurrence network: {B}ased on
  two-mode affiliation network. Physica A: Statistical Mechanics and its
  Applications  \textbf{450},  657--669 (2016)

\bibitem{molontay2019two}
Molontay, R., Nagy, M.: Two {D}ecades of {N}etwork {S}cience as seen through
  the co-authorship network of network scientists. In: International Conference
  on Advances in Social Networks Analysis and Mining, ASONAM. IEEE/ACM (2019)

\bibitem{supp}
Nagy, M., Molontay, R.: Twenty years of network science -- {S}upplementary
  material (2020),
  \url{https://github.com/marcessz/Twenty-Years-of-Network-Science}

\bibitem{newman2018networks}
Newman, M.: Networks. Oxford University Press (2018)

\bibitem{newman2001structure}
Newman, M.E.: The structure of scientific collaboration networks. Proceedings
  of the National Academy of Sciences  \textbf{98}(2),  404--409 (2001)

\bibitem{newman2004coauthorship}
Newman, M.E.: Coauthorship networks and patterns of scientific collaboration.
  Proceedings of the National Academy of Sciences  \textbf{101}(suppl 1),
  5200--5205 (2004)

\bibitem{newman2006finding}
Newman, M.E.: Finding community structure in networks using the eigenvectors of
  matrices. Physical Rev. E  \textbf{74}(3),  036104 (2006)

\bibitem{newman2004finding}
Newman, M.E., Girvan, M.: Finding and evaluating community structure in
  networks. Physical Rev. E  \textbf{69}(2),  026113 (2004)

\bibitem{pawar2019codd}
Pawar, R.S., Sobhgol, S., Durand, G.C., Pinnecke, M., Broneske, D., Saake, G.:
  Codd's world: {T}opics and their evolution in the database community
  publication graph. In: Grundlagen von Datenbanken. pp. 74--81 (2019)

\bibitem{su2010mapping}
Su, H.N., Lee, P.C.: Mapping knowledge structure by keyword co-occurrence: a
  first look at journal papers in {T}echnology {F}oresight. Scientometrics
  \textbf{85}(1),  65--79 (2010)

\bibitem{connected2008}
T\'alas, A.: Connected: {T}he power of six degrees (2008)

\bibitem{uddin2015framework}
Uddin, S., Khan, A., Baur, L.A.: A framework to explore the knowledge structure
  of multidisciplinary research fields. PloS one  \textbf{10}(4),  e0123537
  (2015)

\bibitem{van2009software}
Van~Eck, N., Waltman, L.: Software survey: {VOS}viewer, a computer program for
  bibliometric mapping. Scientometrics  \textbf{84}(2),  523--538 (2009)

\bibitem{vespignani2018twenty}
Vespignani, A.: Twenty years of network science. Nature  \textbf{558},
  528--529 (2018)

\bibitem{watts2004six}
Watts, D.J.: Six degrees: {T}he science of a connected age. WW Norton \& Co.
  (2004)

\bibitem{watts1998collective}
Watts, D.J., Strogatz, S.H.: Collective dynamics of ‘small-world’ networks.
  Nature  \textbf{393}(6684), ~440 (1998)

\bibitem{xia2017big}
Xia, F., Wang, W., Bekele, T.M., Liu, H.: Big scholarly data: {A} survey. IEEE
  Transactions on Big Data  \textbf{3}(1),  18--35 (2017)

\bibitem{yan2014scholarly}
Yan, E., Ding, Y.: Scholarly networks analysis. Encyclopedia of Social Network
  Analysis and Mining pp. 1643--1651 (2014)

\end{thebibliography}
\end{document}